\documentclass[superscriptaddress,titlepage,amsmath,amssymb,aps,pre,reprint]{revtex4-1}
\usepackage{graphicx}
\usepackage{float}
\usepackage[caption=false]{subfig}
\usepackage{tensor}
\newcommand{\mathsym}[1]{{}}
\usepackage{commath}
\newcommand{\unicode}[1]{{}}
\usepackage[usenames]{color}

\begin{document}

\title{How a virus circumvents energy barriers to form symmetric shells}

\author{Sanaz Panahandeh}
\author{Siyu Li}%
\affiliation{%
 Department of Physics and Astronomy, University of California, Riverside, California 92521, USA
}%
\author{Laurent Marichal}%
\affiliation{%
 Universit\'e Paris-Saclay, CNRS, Laboratoire de Physique des Solides, 91405, Orsay, France
}%
\author{Rafael Leite Rubim}%
\affiliation{%
 Universit\'e Paris-Saclay, CNRS, Laboratoire de Physique des Solides, 91405, Orsay, France
}%
\author{Guillaume Tresset}%
\affiliation{%
 Universit\'e Paris-Saclay, CNRS, Laboratoire de Physique des Solides, 91405, Orsay, France
}%
\author{Roya Zandi}%
\affiliation{%
 Department of Physics and Astronomy, University of California, Riverside, California 92521, USA
}%





\begin{abstract}
Previous self-assembly experiments on a model icosahedral plant virus have shown that, under physiological conditions, capsid proteins initially bind to the genome through an en masse mechanism and form nucleoprotein complexes in a disordered state, which raises the questions as to how virions are assembled into a highly ordered structure in the host cell. Using small-angle X-ray scattering, we find out that a disorder-order transition occurs under physiological conditions upon an increase in capsid protein concentrations. Our cryo-transmission electron microscopy reveals closed spherical shells containing \textit{in vitro} transcribed viral RNA even at pH 7.5, in marked contrast with the previous observations. We use Monte Carlo simulations to explain this disorder-order transition and find that, as the shell grows, the structures of disordered intermediates in which the distribution of pentamers does not belong to the icosahedral subgroups become energetically so unfavorable that the caps can easily dissociate and reassemble overcoming the energy barriers for the formation of perfect icosahedral shells. In addition, we monitor the growth of capsids under the condition that the nucleation and growth is the dominant pathway and show that the key for the disorder-order transition in both en masse and nucleation and growth pathways lies in the strength of elastic energy compared to the other forces in the system including protein-protein interactions and the chemical potential of free subunits.  Our findings explain, at least in part, why perfect virions with icosahedral order form under different conditions including physiological ones.
\end{abstract}

\maketitle


\section{Introduction}
The process of formation of virus particles in which the protein subunits encapsidate genome (RNA or DNA) to form a stable, protective shell called the capsid is an essential step in the viral life cycle \cite{Flint,CASPAR1962,nature2016}. The capsid proteins of many small single-stranded (ss)RNA viruses spontaneously package their wild-type (wt) and other negatively charged polyelectrolytes, a process basically driven by the electrostatic interaction between positively charged protein subunits and negatively charged cargo \cite{Sun2007,Siber2008,Gonca2014,Li2017a,asor2019assembly,Paul:13a}. Understanding the phenomena of formation of viral particles is of great interest due to their potential applications in nanomedicine and biomaterials. Capsids can be employed as nanocontainers, biosensors, nanoreactors, and drug or gene delivery vehicles to name just a few \cite{Gonca2016,vernizzi2011platonic,Zlotnick,elife,mohajerani2018role}.

Regardless of the virion size and assembly procedures, most spherical viruses adopt structures with icosahedral symmetry \cite{weiss2005armor,johnson1997quasi,Stefan}.  The total number of proteins in an icosahedral capsid is equal to $60 T$,  where $T$ is the triangulation number assuming only certain integers (T=1,3,4,7...)\cite{caspar1962physical,twarock2019structural}. Independently of the number of proteins, there are always 12 pentamers in a viral shell sitting on the vertices of an icosahedron to preserve the symmetry of the capsid.

How exactly capsid proteins (CPs) assemble to assume a specific size and symmetry have been investigated for over half a century now \cite{Bancroft,Chen:2007b}. Since the self-assembly of virus particles involves a wide range of thermodynamics parameters, different time scales and an extraordinary number of possible pathways, the kinetics of assembly has remained elusive, linked to Levinthal's paradox for protein folding \cite{asor2019assembly,levinthal1969fold,dykeman2014solving,ZANDIreview}. The role of the genome on the assembly pathways and the structure of the capsid is even more intriguing \cite{kindt,Rob1,EricMay,Ben-Shaul2015,Siber2008,LucaRudi2015,GoncaPRL2017}. The kinetics of virus growth in the presence of RNA is at least 3 orders of magnitude faster than that of empty capsid assembly, indicating that the mechanism of assembly of CPs around RNA might be quite different. Some questions then naturally arise: What is the role of RNA in the assembly process, and by what means then does RNA preserve assembly accuracy at fast assembly speed? 

Two different mechanisms for the role of the genome have been proposed: (i) en masse assembly and (ii) nucleation and growth \cite{Mcpherson2001,Hagan2013,perlmutter2014pathways}. Several years ago, McPherson suggested the en masse model in which the nucleic acid attracts CPs in solution to its surface through long-range electrostatic interactions. Note that the assembly interfaces in many CPs are principally short-ranged hydrophobic in character, whereas there is a strong electrostatic, nonspecific long-ranged interaction between RNA and CPs.  To this end, the positively charged domains of CPs associate with the negatively charged RNA quite fast and form an amorphous complex. Hydrophobic interfaces then start to associate, which leads to the assembly of a perfect icosahedral shell. Based on the en masse mechanism, the assembly pathways correspond to situations in which intermediates are predominantly disordered.

More recently Chevreuil {\it et al.} \cite{Chevreuil2018} studied en masse assembly by carrying out time-resolved small-angle X-ray scattering experiments on cowpea chlorotic mottle virus (CCMV), a $T=3$ single-stranded RNA plant virus. 
They found that at neutral pH,
a considerable number of CPs were rapidly ($\sim$28 ms) adsorbed to the genome, which more slowly ($\sim$48 s) self-organized into compact but amorphous nucleoprotein complexes (NPC) (see also \cite{Garmann2016}).
By lowering the pH, they observed a disorder-order transition as the protein-protein interaction became strong enough to close up the capsid and to overcome the high energy barrier ($\sim20k_BT$, where $k_B$ is the Boltzmann constant and $T$ is the temperature) separating NPCs from virions. It is important to note that a marked difference between {\it in vitro} and {\it in vivo} assemblies is that capsid proteins form ordered icosahedral structure at physiological pH ({\it in vivo} conditions) but until now, ordered structures are observed only at acidic pH in the {\it in vitro} self-assembly studies \cite{garmann2014assembly,cadena2012self,DanielDragnea2010,Hsiang-Ku}. 

While because of the electrostatic interaction between the genome and CPs, the en masse assembly is expected to be the dominant assembly pathway, a set of more recent experiments, however, points to the nucleation-elongation mechanism.  Garmann {\it et al.} have employed interferometric scattering microscopy to measure the assembly kinetics of single MS2 virus particles around MS2 RNA strands tethered to the surface of a coverslip \cite{Garmann201909223,DanielDragnea2010}.  Comparisons of individual assembly pathways indicated that most trajectories exhibit a sigmoidal time dependence, with plateau scattering intensity values similar to those acquired from complete particles. According to their experimental data, each trajectory is characterized by a lag time after which the assembly takes off, a signature of the nucleation-elongation mechanism. The measurements of Garmann {\it et al.} reveal that the lag time is not due to the diffusion of the CPs but is directly related to the nucleation and growth mechanism \cite{Garmann201909223}. 

While the experiments of Chevreuil {\it et al.} point to the en masse assembly mechanism, it is hard   to explain how a transition from a disordered amorphous RN-CPs complex to a highly ordered icosahedral capsid can take place where the location of pentamers is very precise and CP-CP interaction is around -7 $k_BT$. Further, we emphasize that the assembly through a nucleation and growth mechanism is not bound to follow an ordered pathway at all times; that is a disorder-order transition might also occur as the shell grows.  The molecular dynamic simulations of Elrad and Hagan have shed light on the assembly pathways of virus particles in the presence of the genome \cite{Hagan2013,Hagan2010}; however, the simulations were designed such that only $T=1$ structures could form whereas the kinetic pathways to the formation of slightly larger viruses such as $T=3$ and $4$ are still elusive. 

In this paper, we study the kinetic pathways of virus assembly using MC simulations and specifically elucidate the role of elastic energy in the disorder-order transition in $T=3$ viruses formed through either en masse or nucleation and growth mechanism. 
We, in particular focus on the impact of protein concentration on the disorder-order transition. Through MC simulations and a set of new experiments employing small-angle X-ray scattering and cryo-transmission electron microscopy, we explore the role of protein concentrations in the formation of $T=3$ structures of CCMV particles. Our findings shed light on the {\it in vivo} assembly of icosahedral structures and can explain to some extent why virion forms at neutral pH {\it in vivo} while the acidic pH was so far required in the {\it in vitro} self-assembly studies of $T=3$ structures.


The paper is organized as follows. In the next section we introduce the simulation and experimental methods. In Sect. III we present the results of numerical simulations indicating the importance of elastic energy in the disorder-order transition. We also study the impact of protein concentration on the assembly pathway and compare it with our SAXS experiments.  Finally, we present our conclusion and summarize our findings. Several details of calculations and
experimental methods are relegated to the Appendix.

\section{Methods}

\subsection{Simulations}

\subsubsection{Viral shell}

To study the growth of viral shells, we use a coarse-grained model in which we focus on the fact that viral capsids are formed basically from hexamers and pentamers. The simplest building blocks that have been extensively used to model viruses and other protein nanocages are trimers \cite{vernizzi2011platonic,SiberZandi2010,rotskoff2018robust}. The model allows us to discretize the growing elastic shell with triangular subunits and associate a monomer with each triangle vertex \cite{wagner2015robust}. 

The total energy of a growing shell is the sum of the attractive interaction between subunits $E_{hp}$, and the stretching $E_s$ and bending $E_b$ energies as follows,
\begin{eqnarray}
  E_s&=&\sum_{l_i} \frac{1}{2}k_s(b_{l_i}-b_0)^2 \label{eqEs}\\ 
  E_b&=&\sum_{<t_i,t_j>}k_b[1-\cos(\theta_{<t_i,t_j>}-\theta_0)]\label{eqEb}\\
  E_{hp}&=&\sum_{v_i} \epsilon_{hp}[nt_{v_i} \cdot (nt_{v_i}-1)], \label{eqEhp}
\end{eqnarray}
where the stretching energy sums over all edges $l_i$ with $b_0$ the equilibrium length and $k_s$ the stretching modulus. The bending energy involves the dihedral angle between connected trimers $<t_i,t_j>$, with $k_b$ the bending rigidity and $\theta_0=2 \arcsin(\frac{b_0}{\sqrt{12 R_0^2-3 b_0^2}})$ the preferred dihedral angle related to the spontaneous radius $R_0$ \cite{wagner2015robust}. The attractive energy due to the hydrophobic interaction sums over all vertices $v_i$, with $\epsilon_{hp}$ being the strength of the monomer-monomer interaction. The range of hydrophobic interaction is considered to be small so that each vertex only interacts with its nearest neighbor vertices. The number of connected trimers is denoted by $nt_{v_i}$ for any vertex $v_i$.  We note that the interaction between subunits is due to both the hydrophobic interactions and electrostatic repulsion between subunits \cite{Kegel2004}. Since the resultant force is attractive for the viruses to assemble, for brevity, we call it the hydrophobic interaction in the rest of the paper.

 Using Monte Carlo (MC) simulations, we then monitor the growth of a viral shell through reversible steps \cite{rotskoff2018robust}.  We consider three different MC moves: 1. diffusion, 2. attachment or detachment, and 3. merging or disjointing. 

\begin{figure}
\centering
\includegraphics[width=0.8\linewidth]{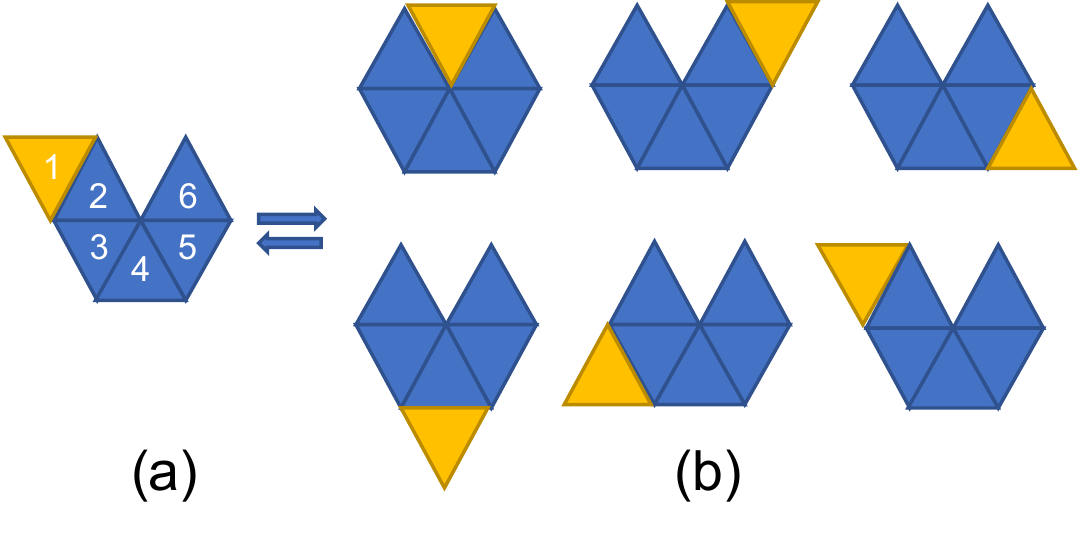}
\caption{Schematic presentation of diffusion of a trimer around the shell edge. (a) Trimer 1 (yellow) is chosen randomly to diffuse around the edge. In this case, only trimers 1 and 6 are allowed to move during the diffusion mode. Trimers 2-5 are connected such that a few bonds need to be broken before they can diffuse. (b) Possible locations for trimer 1 to diffuse.}
\label{trimerdiffuse}
\end{figure}

\begin{figure}
\centering
\includegraphics[width=0.5\linewidth]{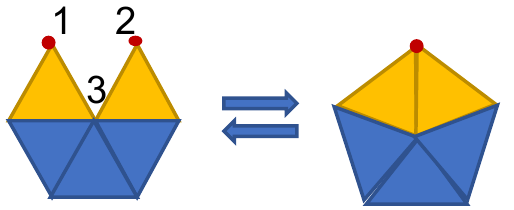}
\caption{Schematic of vertex diffusion: vertices 1 and 2 merge to form a pentamer. The action is reversible in that both merging (left to right arrow) and disjointing (right to left arrow) are possible in the Monte Carlo moves.}
\label{vertexdiffuse}
\end{figure}

During the diffusion process, the number of capsid subunits remains fixed, corresponding to a canonical system.  At each MC step, either a trimer or a vertex is chosen randomly (see Figs.~\ref{trimerdiffuse} and \ref{vertexdiffuse}). The move is accepted with the probability $p_m=\min(1,e^{-\beta E})$ with $\beta=1/k_BT$ and $E$ is the change in the total energy of the system after relaxation. The diffusion of trimers in Fig.~\ref{trimerdiffuse} could be considered as a combination of detachment and growth for trimers, the result of which should be qualitatively the same as diffusion in the canonical ensemble. The diffusion stops based on the convergence of the energy of the system. The arrows in Fig.~\ref{vertexdiffuse} indicate the reversibility of each action.


After the diffusion step, a subunit will be added or removed with the probability $p_g$ and $p_d$, respectively. The probability of growth follows the detailed balance of the grand canonical ensemble with $p_g=\min(1,\frac{T_{g}}{T_{d}}e^{\beta \mu-\beta \Delta E})$ and the probability of removal is $p_d=\min(1,\frac{T_{d}}{T_{g}}e^{-\beta \mu-\beta \Delta E})$\cite{frenkel2001understanding}.  The possible growth and detachment sites are denoted by $T_{g}$ and $T_{d}$ respectively.

The above MC moves can be simply illustrated by the following equations,
\begin{eqnarray}
C & \rightleftharpoons & C' \label{diffuse}\\
C + T & \rightleftharpoons & TC \label{grow}
\end{eqnarray}
where Eq.~\ref{diffuse} involves the trimer diffusion, vertex moves, two trimers merging or disjointing while the total number of triangles are kept constant. Equation ~\ref{grow} corresponds to the shell growth or dissociation where a trimer joins to or detaches from the capsid.

We note that by monitoring the assembly of proteins around the genome in most experiments, RNA condenses into its final size much faster than the other time scales involved in the process \cite{Borodavka2012,Chevreuil2018,DanielDragnea2010}. Thus, we replace the genome with a spherical core as the goal of this paper is basically on how the shell overcomes many energy barriers to form symmetric structures and not on the genome configurations. Replacing the genome by a spherical core will decrease many degrees of freedom in the system, allowing us to focus carefully on the energetics involved in the assembly of viral shells. Thus, we consider a spherical core interacting through Lennard-Jones (LJ) with the capsid proteins $E_{lj}=\sum_{vi} \epsilon_{lj} [(\frac{\sigma_{lj}}{r_{<vi,g>}})^{12}-2(\frac{\sigma_{lj}}{r_{<vi,g>}})^6 ]$
, where $\epsilon_{lj}$ is the potential strength and $r_{<vi,g>}$ is the distance between the core center and the triangle vertex $v_i$.

\subsection{Experiments}

\subsubsection{Sample preparation}
CCMV virions were purified from infected blackeye cowpea leaves (\textit{Vigna ungiculata}) after the protocol developed by Ali and Roossinck \cite{Ali2007}. Briefly, infected leaves were blended with 0.15 M sodium acetate pH 4.8 and ice-cold chloroform prior to precipitation of the virions with 8\% (v.w$^{-1}$) poly(ethylene glycol). The virions were further purified through a 20\% (v.w$^{-1}$) sucrose cushion and stored at -80$ ^\circ$C until use. Proteins were purified from virions via RNA precipitation with 0.5M CaCl$_2$ and stored at 4$ ^\circ$C until use. Full RNA genome was extracted from virions by using TRIzoll$\circledR$ reagent (Life Technologies, France) according to the protocol recommended by the manufacturer. CCMV RNA 2 was transcribed \textit{in vitro} with a linearized plasmid coding for the RNA and containing a T7 promoter. All the RNA were stored in RNase-free water at -80 $^\circ$C. Nucleoprotein complexes (NPCs) were assembled by mixing capsid proteins and RNA in the desired ratio and in some cases, dialyzed against the final buffer. The detailed protocols for sample preparation are available in the Supporting Information (SI).

\subsubsection{Small-angle X-ray scattering (SAXS)}
Scattering patterns were recorded at the ID02 and SWING beamlines of the European Synchrotron Radiation Facility (Grenoble, France) and the SOLEIL synchrotron (Saint-Aubin, France), respectively. Between 10 and 100 frames were collected for each sample with a beam exposure time set to 10 ms. \textit{Ab initio} shape reconstructions were performed by using the programs DAMMIF \cite{Franke2009} and GNOM \cite{Svergun1992}. For each shape reconstruction, 20 models were averaged then filtered, and the final structures were rendered with Chimera \cite{Pettersen2004}. Sphericity indexes were computed through a principal component analysis. Let $a$, $b$ and $c$ be the standard deviations in descending order along the principal axes of a reconstructed structure, the sphericity index is defined as $[c^2/(ab)]^{1/3}$. The sphericity index is positive and tends to one for a sphere.

\subsubsection{Cryo-transmission electron microscopy}
The sample was frozen on a holey carbon grid (Quantifoil R2/2) by using a FEI Vitrobot and observed with a JEOL JEM-2010 electron microscope equipped with a 200-kV field emission gun. Images were collected with a Gatan Ultrascan 4K CCD camera with a $\times$50,000-magnification using a minimal dose system. Defocus was set to 2.5 $\mu$m.

\section{\label{result}Result and Discussion}

As noted in the introduction, two different mechanisms are proposed for the role of the genome in the assembly of viral shells: (1) nucleation and growth and (2) en masse assembly. We will decipher each in the following sections.  Because of the extraordinary number of degrees of freedom involved in the assembly process, we focus only on understanding a number of recent experiments in which the genome has assumed different roles.  We study the experiments of Garmann {\it et al.} \cite{Garmann201909223} to elucidate the process of nucleation and growth in the presence of the genome and then investigate the experiments of Chevreuil {\it et al.} to explore how the genome promotes the assembly through an en masse or a multinucleation mechanism \cite{Chevreuil2018}. In order to obtain a better understanding of the disorder-order transition, we perform a new set of X-ray scattering experiments at neutral pH to study the role of protein concentration in the disorder-order transition and compare them with the results of our simulations.  In what follows, we first present the results of our simulations for the single nucleation assembly and then show our experimental data and simulations corresponding to the multinucleation or en masse mechanism. We emphasize that the main difference between the en masse and nucleation and growth mechanisms corresponds to their relevant time scales. In the en masse process, subunits bind to the genome in a short time frame, forming multiple patches or nuclei, and the structure keeps growing in a disordered manner. When enough subunits are adsorbed, they undergo cooperative rearrangements to assemble into an ordered capsid. By contrast, in the nucleation-elongation mechanism, the longer time scale corresponds to the formation of the first nucleus, and the shell grows by sequential binding of subunits to its edge. The growth phase is always shorter than the nucleation time. The assembly starts at one point and continues until an ordered shell is formed. Both en masse and nucleation-elongation pathways can involve a disorder-order transition.

\subsection{Nucleation and growth}\label{nucleationsection}

We first investigate the assembly of empty capsids where the nucleation and growth is the dominant mechanism and then present our results on the single nucleation mechanism in the presence of genome.

\subsubsection{Empty capsids}

Following Eqs.~\ref{eqEs}-\ref{eqEb}, the total energy of an empty capsid can be written as

\begin{equation}\label{reveq}
E_{tot}=E_{el}+E_{hp}-N_T~\mu=(\epsilon_{el}+\epsilon-\mu)N_T,
\end{equation}
where $N_T$ is the total number of trimers, $E_{el}=E_s+E_b$ is the total elastic energy of the capsid, $E_{hp}$ is the protein-protein interaction (basically hydrophobic), and $\mu$ is connected to the chemical potential of free proteins in solution. For a dilute system, $\mu=k_BT \log \frac{\rho}{\rho_0}$, with $\rho$ being the density of free proteins in solution and $\rho_0$ the reference density. The average elastic energy and attractive protein-protein interaction per subunit are denoted by $\epsilon_{el}$ and $\epsilon$, respectively. 

\begin{figure*}
\centering
\includegraphics[width=\linewidth]{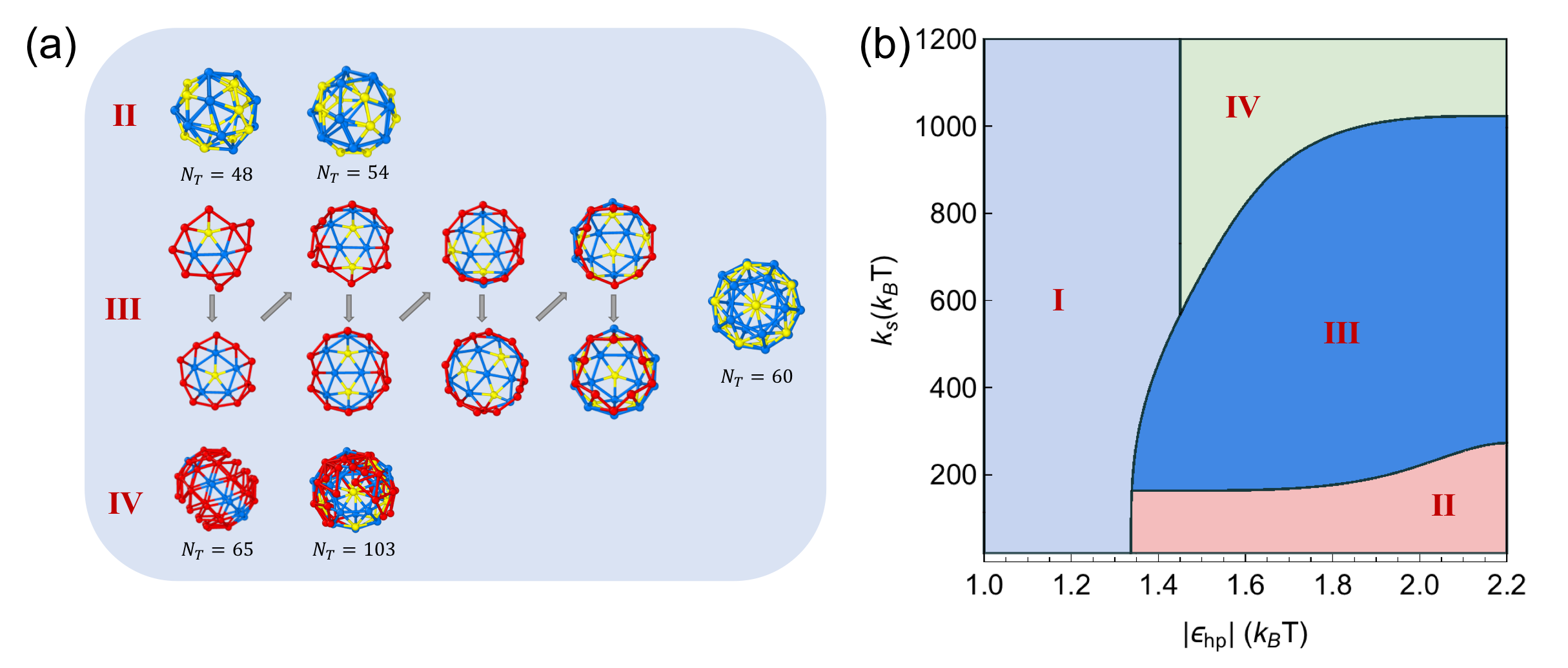}
\caption{Phase diagram for the reversible growth model and snapshots of assembled structures as a function of stretching modulus, $k_s$, and monomer-monomer hydrophobic interaction, $\epsilon_{hp}$. In region I, no shell forms due to lack of nucleation; in region II, capsids close to form irregular shapes with lower symmetries, and total trimer number, $N_T$, is less than 60; in region III, only $T=3$ structures form.  Snapshots of a representative pathway are illustrated in the panel (a) where pentamers first form at the boundary, and then the shell continues to grow around them. Images in panel (a) are made with OVITO\cite{ovito}. The pathway is consistent and matches very well with the ground-state calculation corresponding to the locations of pentamers as a shell grows (see Appendix Fig.~14). In region IV, the elasticity is so high that the structures cannot close but form messy shells. Other parameters of simulations are spontaneous radius $R_0=1.5$, bending rigidity $k_b=200 k_BT$, and chemical potential $\mu=-14.6 k_BT$.}
\label{rev_phase}
\end{figure*}

Figure \ref{rev_phase}a.III illustrates the snapshots of simulations for the growth of an empty shell through the reversible pathway described in the previous section.  As the cap grows, at some point, the formation of pentamers becomes energetically unavoidable because of the spherical geometry of the shell. The first pentamer always forms in the vicinity of the cap edge, and then the shell grows around the pentamer such that, at the end, the pentamer is in the middle of the cap. The following pentamers appear as illustrated in the figure, and the shell grows around them. We note that the hydrophobic interaction always prefers that each subunit has the maximum number of neighbors, six. However, because of the shell curvature, this is not always possible. The stretching energy defines the position
of pentamers and thus the symmetry of the shell.

Quite interestingly the position of pentamers follows very well the prediction of continuum elasticity theory where the ground state energy of a spherical cap was calculated to obtain the optimal position of disclinations \cite{lizandigrason,li2018large}, see Appendix Fig.~14. This confirms that the system is completely reversible, equilibrated, and is able to find the minimum energy structures along the pathway.  

The fact that most spherical viruses adopt structures with icosahedral symmetry reveals the important role of elasticity in the energetics of viral shells \cite{wagner2015robust,li2018large,panahandeh2018equilibrium}.  Nevertheless, it has remained a mystery how an error-free shell formed out of 90 dimers or 60 trimers grows with perfect icosahedral symmetry under many different {\it in vitro} assembly conditions. One expects that when a considerable part of a shell is formed, the pentamers formed well inside the cap far from the edge will become more or less frozen at their locations.  Thus, if a pentamer is formed in a ``wrong'' position, the icosahedral symmetry should permanently be broken. However, the reversible simulations show that, as a shell grows, the elastic energy can become strong enough to easily break the bonds for a wide range of hydrophobic interactions and repair the positions of pentamers formed in the locations not consistent with icosahedral subgroup symmetries. 

\begin{figure}
\centering
\includegraphics[width=1.1\linewidth]{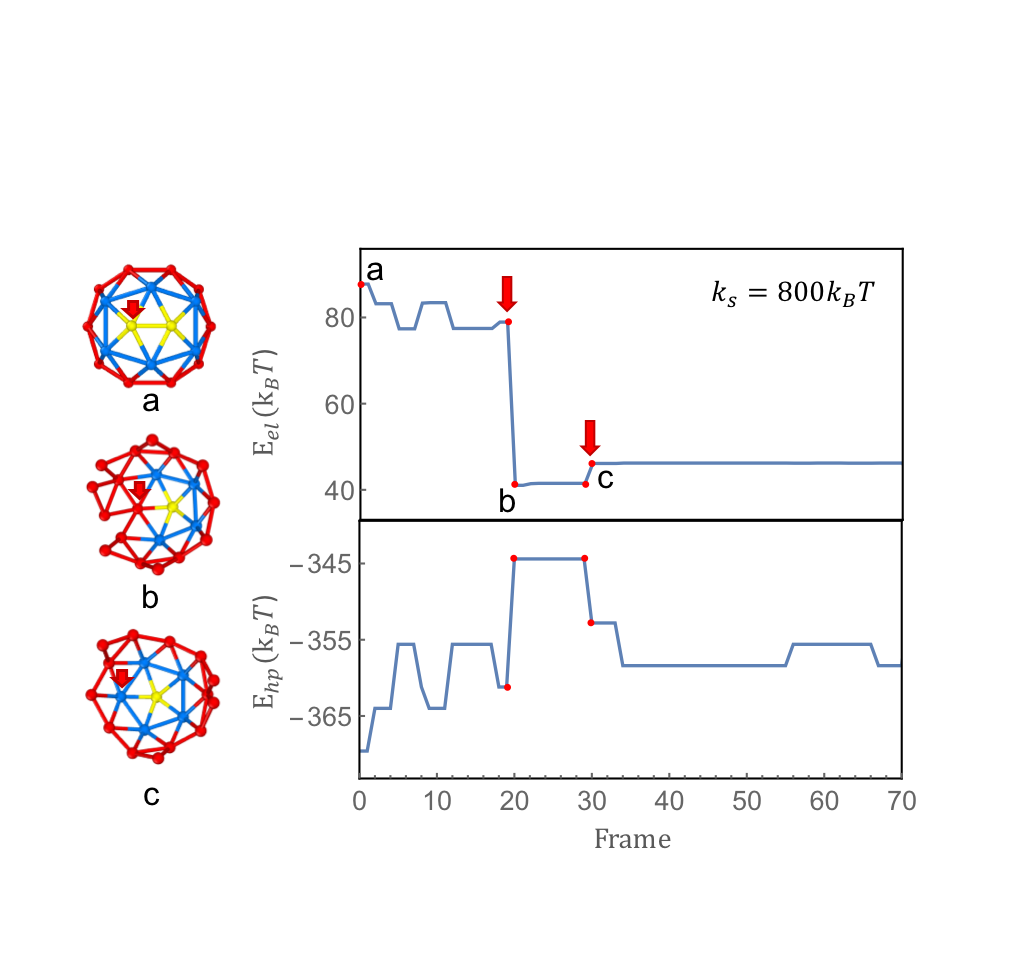}
\caption{Snapshots of simulations for a partially formed capsid with 24 subunits. The right column shows the elastic and hydrophobic energies vs.~frame (number of MC moves) or time. (a) Two pentamers are formed in the vicinity of each other.  (b) Around 20 MC steps due to high elastic energy (marked with a red arrow), a bond breaks to decrease the elastic energy while increasing the hydrophobic energy (point b in the energy plots). (c) Hexamer forms and slightly increases the elastic energy while decreasing the hydrophic energy.  The protein subunits stretching rigidity is $k_s=800 k_BT$ and bending rigidity $k_b=200 k_BT$. Other parameters used are hydrophobic interaction $\epsilon_{hp}=-1.4 k_BT$ and chemical potential $\mu=-14.6 k_BT$.}
\label{penfix}
\end{figure}

Figure~\ref{penfix} shows the snapshots of simulations of a cap in which two pentamers are formed next to each other at the beginning of the assembly.
The reversible growth allows the bonds to break and a pentamer to change to a hexamer as the assembled shell grows to have about 25 trimers. Note that the dynamic of system specifically depends on the stretching modulus of subunits, $k_s$. Lower stretching rigidity slows down the pentamer-hexamer transition, emphasizing the role of elasticity in the kinetic pathway of assembly. Fig.~15 in Appendix shows that as $k_s$ decreases from 800$k_BT$ to 200$k_BT$, it takes much longer for the system to repair the position of the pentamer.
 
 Figure~\ref{penfix} clearly illustrates that the elastic energy of the shell decreases (indicated with an arrow) when the bonds of a ``wrong'' pentamer are broken.  During this time, the hydrophobic energy increases, which is the penalty for breaking the bonds, resulting into an energy barrier to the formation of icosahedral shells. However, in this case, the elastic energy of the shell is strong enough to overcome the local energy barriers and assume the symmetric structure. It is important to emphasize that the elastic energy diverges if a shell grows without pentamers.  
 
If the elastic energy is too small or too large compared to the attractive interaction between subunits, the symmetry of the shell will also be broken.  To explore the interplay of elasticity and the hydrophobic interaction between subunits in the final structure of shells, we construct a phase diagram for the spring modulus $k_s$ ranging from $20k_B T$ to $1200k_B T$ and \abs{\epsilon_{hp}}$\in$ [1$k_BT$,2.2$k_BT$] as illustrated in Fig.~\ref{rev_phase}b.  The value of bending rigidity was chosen based on the experimental values obtained previously \cite{roos2010physical,Zeng2017a}. Note that as long as the ratio of the stretching to bending modulus is larger than 1, the final assembly products do not depend on the bending modulus, which we have kept equal to $k_b = 200k_B T$ for all of the simulations \cite{roos2010physical,Zeng2017a}.

In region I of the phase diagram, the hydrophobic interaction is so weak that no capsid could nucleate.  In region II the elastic energy is weak compared to the hydrophobic interaction, and the shell easily gets kinetically trapped and irregular structures form with total number of subunits less than 60 (a $T=3$ structure)(Fig.~\ref{rev_phase}a). One can see that for a large region of the phase diagram (region III), the competition between elastic and hydrophobic is such that the final product is always a $T=3$ structure.  As the elastic energy becomes stronger in region IV,
even for very strong hydrophobic interactions, the shell is not be able to close but assumes a structure with cylindrical symmetry and grows ``branchy'' way, forming a messy structure (Fig.~\ref{rev_phase}a). While $T=3$ constitutes the minimum energy structure for the entire phase diagram, messy structures with no symmetry can readily form in different regions as shown in Fig.~\ref{rev_phase}. It is not always possible to obtain $T=3$ structures without using different MC techniques to avoid local traps. We calculate the energy barrier to the formation of $T=3$ structures in the next section to explain why in some regions the shells can easily get trapped in a local minimum energy structure.
 
 At the end of this section, we emphasize that even though during the simulations one subunit was added at a time, our conclusion remains the same if several subunits join the growing shell simultaneously. The strength of elastic energy as the shell assembles makes the process reversible and the results robust, independent of the number of subunits joining the growing shell .

\subsubsection{classical nucleation and energy barrier}

To understand why closed structures without any particular symmetry appear in the phase diagram of Fig.~\ref{rev_phase} where $T=3$ structure has the lowest energy, we calculate the energy barriers for the formation of $T=3$ in different regions of the diagram. 

For a complete shell, the total hydrophobic interaction is $E_{hp}=15(Q-4)\epsilon_{hp}$ where Q is the total number of subunits and $\epsilon_{hp}$ is the monomer-monomer interaction as noted before. For a growing shell or a cap, the hydrophobic interaction in Eq.~\ref{eqEhp} becomes

\begin{eqnarray}
\label{partial_hp}
    E_{hp}&=&\left(15N_T-60\frac{N_T}{Q}-9\sqrt{\sqrt{3}\pi}\sqrt{\frac{N_T}{Q}(Q-N_T)}\right)\epsilon_{hp} \nonumber\\
    &=&15\epsilon_{hp} \frac{Q- 4}{Q}N_T-15\epsilon_{hp}\sqrt{\alpha}\sqrt{\frac{N_T}{Q}(Q-N_T)} \nonumber\\
    &=&\epsilon N_T + \epsilon_{l},
\end{eqnarray}
where $\alpha=\frac{9\sqrt{3}\pi}{25}$ (see Section.~3 in Appendix). The average hydrophobic interaction per subunit is $\epsilon=15\epsilon_{hp} \frac{Q-4}{Q}$ and converges to $15\epsilon_{hp}$ as $Q\rightarrow+\infty$. 

Combining Eq.~\ref{reveq} and Eq.~\ref{partial_hp}, we obtain the free energy of the growing capsid as
\begin{eqnarray} \label{totalenergy}
    \frac{E_{tot}}{\lvert \epsilon\rvert}&=&\frac{\epsilon_{el}+\epsilon-\mu}{\lvert \epsilon\rvert}N_T+\sqrt{\alpha}\frac{Q}{Q-4}\sqrt{\frac{N_T}{Q}(Q-N_T)} \nonumber\\
    &=&A N_T + a \sqrt{N_T(Q-N_T)},
\end{eqnarray}
where $a=\frac{\sqrt{\alpha Q}}{Q-4}$ is a geometric factor and $A=\frac{\epsilon_{el}+\epsilon-\mu}{\lvert \epsilon\rvert}$ measures the difference between the chemical potentials of free subunits in solution and in full capsids.  To make it dimensionless, we have divided it by the hydrophobic energy per subunit $\epsilon$.
For a $T=3$ structure 
\begin{equation} \label{Avalue}
    A=\frac{\epsilon_{el}+14\epsilon_{hp}-\mu}{\lvert 14\epsilon_{hp}\rvert}.
\end{equation}  
The height of energy barrier and the nucleation size can then be calculated through the maximum of the free energy as follow:
\begin{eqnarray}
N_T^*&=&\frac{1}{2}Q\left(1+\frac{A}{\sqrt{A^2+a^2}}\right)\nonumber\\
{E_{tot}}^*&=&\frac{1}{2}\lvert \epsilon\rvert Q\left(A+\sqrt{A^2+a^2}\right)
\end{eqnarray}
Or
\begin{eqnarray}
N_T^*&=&\frac{1}{2}Q\left(1+\frac{\Gamma}{\sqrt{\Gamma^2+1}}\right)\nonumber\\
{E_{tot}}^*&=&\frac{1}{2}\lvert \epsilon\rvert a Q\left(\Gamma+\sqrt{\Gamma^2+1}\right),
\end{eqnarray}
where $\Gamma=\frac{A}{a}$ is the supersaturation.

\begin{figure}
    \centering
    \includegraphics[width=0.5\textwidth]{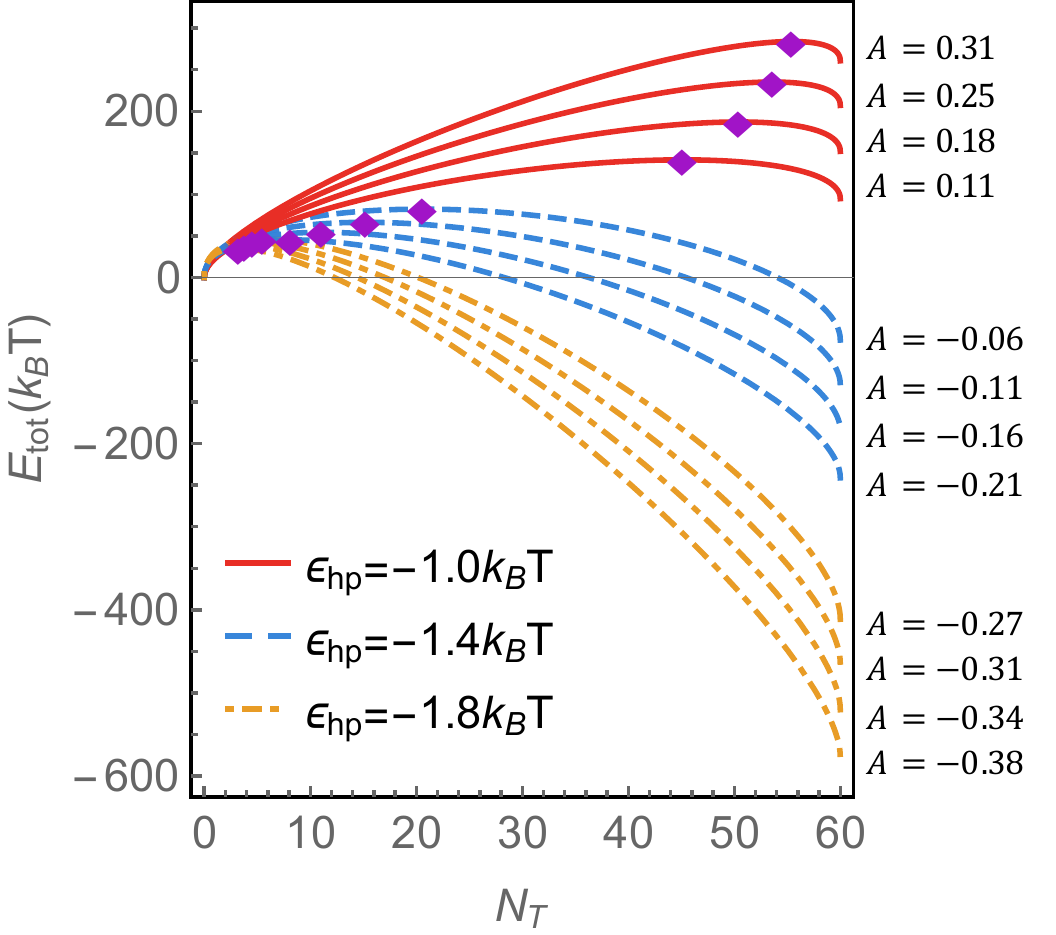}
    \caption{Total energy $E_{tot}$ as a function of subunits. The energy barriers are indicated with diamonds. The value of coefficient $A$ (Eq.~\ref{Avalue}) is indicated next to each curve. The stretching moduli used are $k_s=800,600,400$ and $200k_B T$~(from top to bottom for each color). The chemical potential is $\mu=-14.6k_B T$, and the hydrophobic interaction is $\epsilon_{hp}=-1.0k_B T~(\text{solid})$, -1.4$k_B T$~(\text{dashed}), and -1.8$k_B T$~(dotted dashed). 
    }
    \label{fig:barrier}
\end{figure}

Figure~\ref{fig:barrier} shows the total free energy (Eq.~\ref{totalenergy}) vs. the number of subunits for $A\in (-0.5,0.5)$ (Eq.~\ref{Avalue}) at three different values of the monomer-monomer hydrophobic interaction, $\epsilon_{hp}=-1.0,-1.4,$ and $-1.8k_B T$. The figure reveals the impact of different thermodynamics parameters on the height of energy barriers.  As illustrated in the figure, for a fixed $\mu=-14.6k_BT$ and a given hydrophobic interaction $\epsilon_{hp}$, the energy barrier increases when the elastic energy per subunit $\epsilon_{el}$ increases (see Eq.~\ref{Avalue}). 

The results shown in Fig.~\ref{fig:barrier} are consistent with the phase diagram presented in Fig.~\ref{rev_phase} in that for $\mu=-14.6k_BT$ if factor $A$ is large because of the strong elastic energy and the small hydrophobic $\epsilon_{hp}=-1k_BT$, then the energy barrier is too high for the formation of capsids and no structure nucleates.
Note that if the energy of the final product is positive, the
symmetric shells are thermodynamically unstable and no
capsids form. In fact, for $\epsilon_{hp}=-1.0k_BT$ regardless of stretching modulus, no capsid nucleates unless the chemical potential is at least $\mu=-10k_BT$.   On the other hand, when hydrophobic interaction is too strong compared to the elastic energy, $A$ is small and the nucleation size is less than five subunits, resulting in the formation of kinetically trapped smaller structures than $T=3$, corresponding to region II in the phase diagram.  

In the next section, we present the growth of capsids in the presence of genome through a nucleation and growth mechanism. 
\subsubsection{Assembly in the presence of genome}

The self-assembly experiments of Garmann {\it et al.} indicate the assembly proceeds through a nucleation and growth mechanism with genome present \cite{Garmann201909223}.  The process of nucleation and growth on the surface of a gold nanoparticle or condensed genome is very similar to that of an empty capsid. In the former case, we start the simulations assuming that one trimer is sitting on the core interacting attractively with it.  
 
Mimicking the experiments of Garmann {\it et al.} we assume that the concentration of proteins is such that the process is not diffusion limited; that is, there are enough proteins in the vicinity of genomes to form capsids.  More specifically, we consider that the subunits diffuse from the solution to the core, but the duration of attachment of a single protein to the core is smaller than the nucleation time.  Thus, the delay in the assembly is just because of nucleation. 
To this end, we only focus on the assembly of proteins on the core and do not study the diffusion of proteins to the core.  The total energy of the complexes of core and proteins can be written as
\begin{equation}\label{totE}
E_{tot}=E_{lj}+E_{el}+E_{hp}-N_T~\mu=(-\epsilon_{lj}+\epsilon_{el}+\epsilon-\mu)N_T,
\end{equation}
where $\epsilon_{lj}$ is the depth of the LJ potential representing the core-protein interaction on the core surface. For simplicity, we always set $R_{core}=R_0$.

 The results of simulations in the presence of core are presented in Fig.~\ref{timescalesingle}. For these sets of simulations, we have used the experiments of Garmann {\it et al.} to calibrate our MC steps to seconds \cite{Garmann201909223}. During the simulations, we observe that after about four or five subunits are assembled, the growth becomes very fast. As shown in the figure, it takes about 70 s on average for the $\mu=-10.5k_BT$ (Orange lines) capsid to nucleate, which is followed by a rapid growth phase of 33 s. 
 As the protein concentration goes higher ($\mu=-10k_BT$, blue lines), the nucleation time becomes shorter, about 24 s, and the rapid growth phase becomes about 25 s during which the capsid assembles to almost its final size. Note that, in these cases, in the absence of genome, no capsids could form. Changing $\mu$ from -10.5$k_BT$ to -10$k_BT$, we increase the protein concentration by 64\% and observe that the nucleation time decreases about 65\%.  This is consistent at least qualitatively with the results of Gramann {\it et al.}\cite{Garmann201909223}, which according to their experiments, as the protein concentration changes from 1.5 to 2$\mu$M, the nucleation time decreases about from about 160 to about 95 s. The results of our simulations also show that the growth time decreases with increasing protein concentration, but at a much slower rate than the nucleation time, as seen in the experiments. 
 
 If we increase the protein concentration even further to $\mu=-8k_BT$, the chance of formation of another nucleus will be high and the process might become multinucleus or en masse assembly, as will be explained in the next section.  All of these simulations show that the interaction of capsid proteins with genome facilitates the formation of capsids; that is, it lowers the energy barrier. 
 

The above results show that, in the presence of genome,
even if the protein-protein interaction is not strong enough for the formation of empty capsids and the protein-genome interaction is weak such that the rate of detachment is larger than the rate of capsid nucleation, the combined effect of both interactions can give rise to the assembly of filled capsids.  Note that if the presence of the genome decreases the energy barrier only by 1 or 2$k_BT$, because of the associated Boltzmann factor, the rate of formation of the capsid will increase by a factor of 3 or 4.
 
In the next section, we will study the en masse or multinucleation mechanism. First, we present our new experiments and then the results of our simulations assuming that the protein-genome interaction is so strong that the proteins can easily get adsorbed to the genome and then nucleation occurs in different positions on the core.
 
\subsection{En masse assembly}\label{enmasssection}
  
 \subsubsection{Experimental evidence of a disorder-order transition at neutral pH}
  
The kinetic studies of Chevreuil {\it et al.} \cite{Chevreuil2018} performed by mixing CCMV subunits (dimers of CPs) and the full (single-stranded RNA) genome at neutral pH showed that the assembly proceeded through an en masse rather than through a single nucleation and growth process, and the final objects were amorphous nucleoprotein complexes (NPCs). Quite intriguingly, when the genome was replaced by a flexible, linear polyelectrolyte, namely, poly(stryrene sulfonic acid), \cite{Hu2008,Tresset2014} small but closed structures were formed, again via the en masse pathway \cite{Chevreuil2018}. Why then cannot subunits form ordered structures in the presence of genomic RNA at neutral pH, which is yet physiologically relevant?

\begin{figure}
\centering
\includegraphics{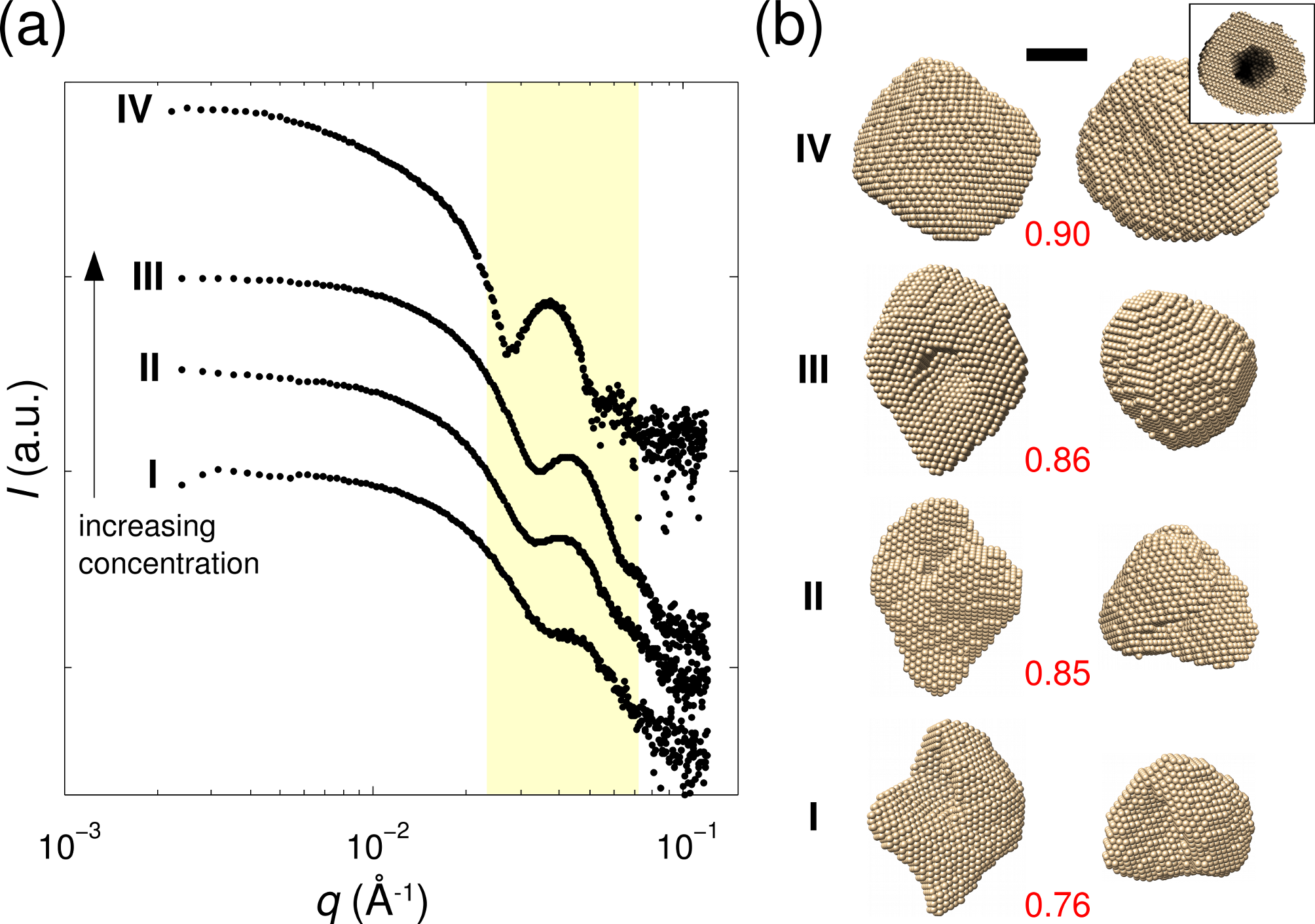}
\caption{Equilibrium structure of NPCs at neutral pH and increasing concentrations in subunits and genomic RNA. (a) SAXS patterns obtained at a fixed subunit-to-RNA mass ratio of 6. Samples \textbf{I}, \textbf{II}, and \textbf{III} are made with the full CCMV genome at subunit concentrations of 0.5, 1.0, and 2.1 g.L$^{-1}$, respectively, whereas sample \textbf{IV} contains only \textit{in vitro} transcribed CCMV RNA 2 for a subunit concentration of 2.1 g.L$^{-1}$. The yellow area highlights the growth of an interference peak in the curves due to an increasingly well-defined length scale in the scattering objects. The scattering curves are shifted for clarity. (b) \textit{Ab initio} shape reconstructions carried out with the scattering curves in (a). The numbers in red are the sphericity indexes calculated for the corresponding structures. The inset is a cross-sectional view of the structure obtained for sample \textbf{IV}. Scale bar is 10 nm.}
\label{fig:npc_fit_dammif}
\end{figure}

Figure~\ref{fig:npc_fit_dammif} illustrates the effect of an increase in the concentration of subunits and genomic RNA on the structure of NPCs at neutral pH. SAXS patterns of NPCs (Fig.~\ref{fig:npc_fit_dammif}a) made with the full CCMV genome exhibit an increasingly pronounced interference peak in the medium $q$ values as the concentrations are increased. The peak is even sharper when the full genome is replaced solely by \textit{in vitro} transcribed RNA 2. The reconstructed shapes and their associated sphericity indexes (Fig.~\ref{fig:npc_fit_dammif}b) suggest that the structures become more ordered and gradually acquire a spherical symmetry at high concentrations. Notice that the reconstructed shape of sample \textbf{IV} displays a hollow core (see inset of Fig.~\ref{fig:npc_fit_dammif}b) like native virions \cite{Fox1998}. The reconstructed shapes must be considered with caution though, as they only provide ensemble-averaged, low-resolution structures and cannot reflect the actual variability between individual objects. Note that the analysis of the SAXS patterns by a polydisperse vesicle model (see Appendix Fig.~17) also supports a higher degree of order at high concentrations. 

\begin{figure}
\centering
\includegraphics[width=\linewidth]{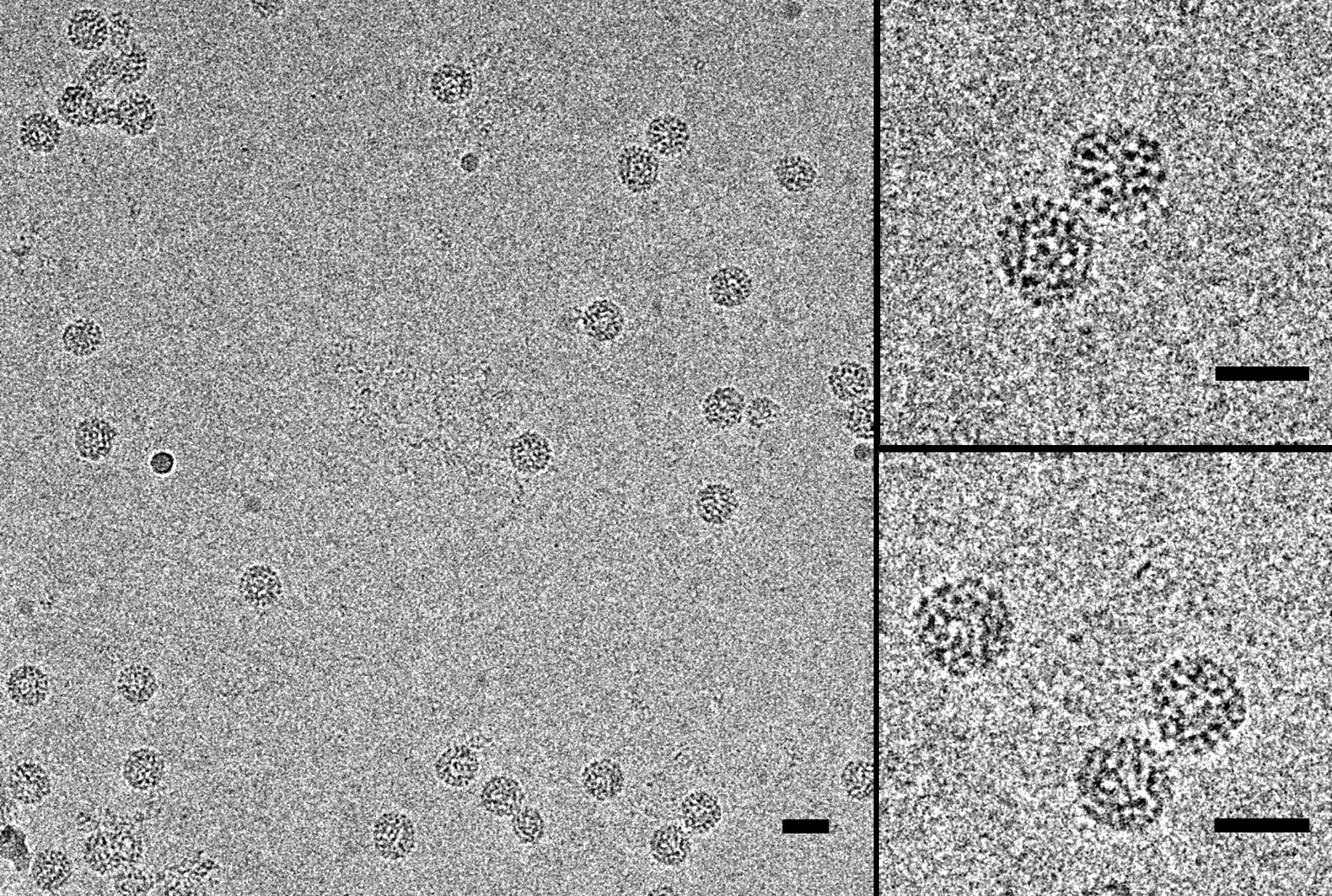}
\caption{Cryo-transmission electron microscopy images of NPCs prepared at neutral pH. CCMV subunits are at a concentration of 3.9 g.L$^{-1}$, and \textit{in vitro} transcribed CCMV RNA 2 is mixed at a subunit-to-RNA mass ratio of 6. Scale bars are 30 nm.}
\label{fig:npc_rna2_high_conc_cryotem}
\end{figure}

\begin{figure}
\centering
\includegraphics[width=0.5\textwidth]{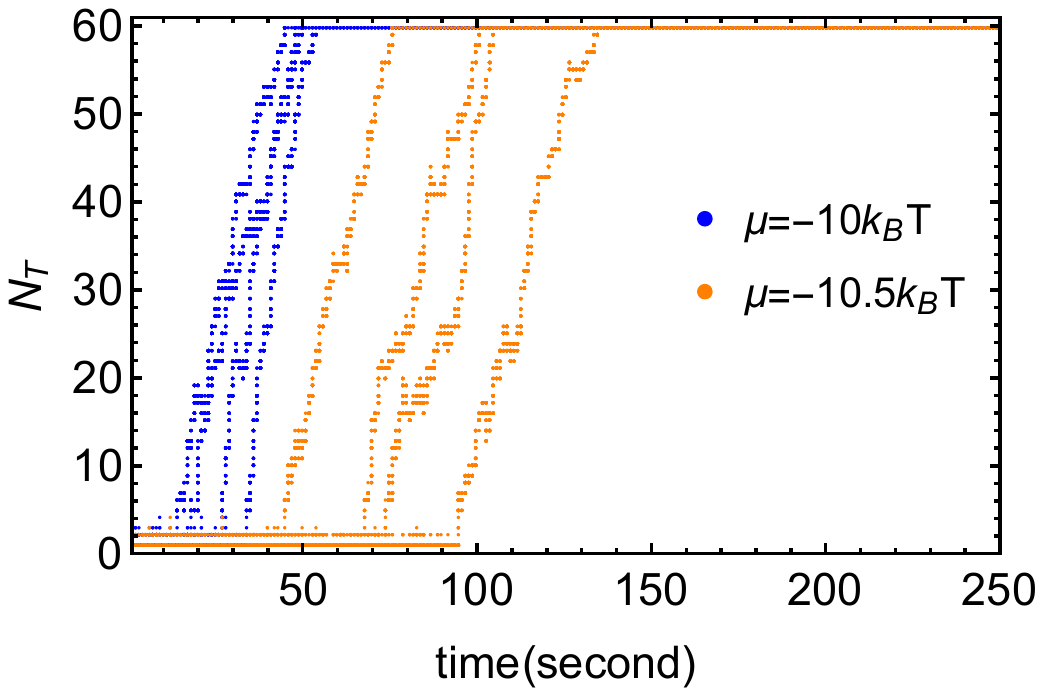}
\caption{Several pathways for the formation of $T=3$ structures. The plot shows the number of subunits in the assembled caps as a function of time (second) for two different of chemical potentials $\mu=-10$ $k_BT$(Blue color) and $\mu=-10.5k_BT$(Orange color).  For each color, four different runs are shown revealing fluctuations in the pathways of formation of $T=3$ icosahedral structures. The other parameters in simulations are $k_s$=600$k_BT$, $k_b$=200$k_BT$, $\epsilon_{hp}$=-1$k_BT$ and the genome-protein interaction $\epsilon_{lj}$=1$k_BT$. For these parameters, the assembly only proceeds through the nucleation and growth mechanism.
}
\label{timescalesingle}
\end{figure}

In order to get a better insight into the morphology of the assembled objects, we perform cryo-transmission electron microscopy on a highly concentrated sample of NPCs made with \textit{in vitro} transcribed RNA 2 because the SAXS patterns reveal objects with the highest sphericity index (Fig.~\ref{fig:npc_fit_dammif}b). Quite consistently, the imaged objects (Fig.~\ref{fig:npc_rna2_high_conc_cryotem}) are mostly ordered and morphologically similar to native virions in these conditions, with a diameter around 30 nm. Some objects are still slightly disordered or aggregated. No empty spherical capsids are found but we can see a few hollow nanotubes (see Appendix Fig.~18) in agreement with the phase diagram of CCMV CPs at neutral pH \cite{Lavelle2009}.

In summary, our experimental studies show that a disorder-order transition occurs within NPCs upon an increase in subunits and genomic RNA concentrations, and that closed spherical shells packaging viral RNA can be assembled at neutral pH. These findings motivate us now to investigate theoretically the ability of the chemical potential to lower the energy barrier between NPCs and assembled virions.

\subsubsection{Simulations}

\begin{figure}

\centering
\includegraphics[width=0.5\textwidth]{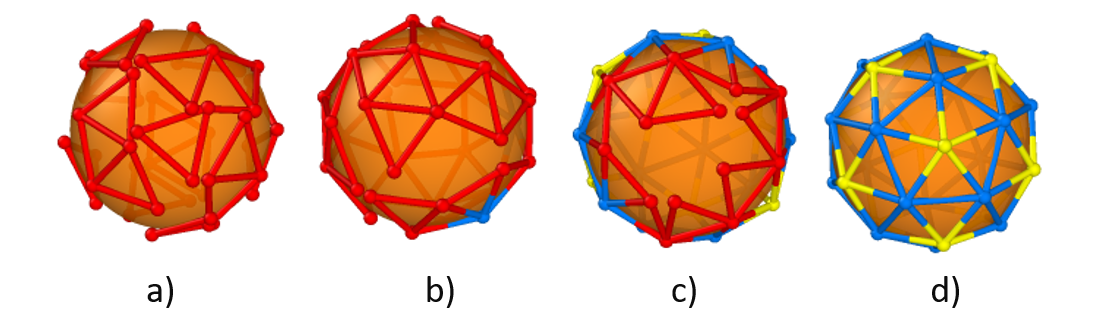}
\caption{Snapshots of simulations in Fig.~\ref{timescalemulti}.  The chemical potential is $\mu=-14k_BT$, $k_s$=600$k_BT$, $k_b$=200$k_BT$ and $\epsilon_{lj}=11.2k_BT$. (a) View of an amorphous nucleoprotein complex after less than a second for $\epsilon_{hp}=-0.7k_BT$.  There are around 20 subunits on the surface of the core.(b) View of a disordered structure. After 96 s there are 40 trimers on the genome and multiple nuclei have formed. Later when the capsid has 45 trimers, we increase the hydrophobic strenght by changing $\epsilon_{hp}$ from $-0.7k_BT$ to $-1k_BT$. This is consistent with the experiments when the pH is lowered from 7.5 to 5.2. (c) View of a partially formed capsid with 54 trimers after 5000 s. Subunits are relaxed to their lowest minimum energy positions.(d) Structure of a closed $T=3$ capsid with 60 trimers. It takes around 14000 s for the capsid to complete. Red colors corresponds to vertices on the edge. Blue and yellow colors show the position of hexamers and pentamers respectively. The images are made with OVITO\cite{ovito}.}
\label{multinuc}
\end{figure}

The snapshots of the growth of a capsid through en masse is illustrated in Fig.~\ref{multinuc}. Right at the beginning of simulations,  many subunits get adsorbed to the core and a messy structure forms (Fig.~\ref{multinuc}a), similar to the NPCs observed in previous experiments \cite{Garmann2015,Chevreuil2018} at pH 7.5 and the new ones presented in the last section. Special care is taken so that there is no overlap between subunits.  We note that as for the simulations in which the nucleation and growth is the dominant mechanism, we do not study the diffusion of proteins into the genome in the case of en masse (or multinucleations) assembly either. More specifically, we consider the assembly is reaction-limited, and thus the important time scale in the problem involves the addition of subunits to each other in order to form larger patches.  Due to strong protein-genome interaction,  we start simulations with about 10-15 subunits attached to the genome and consider that the growth proceeds basically through joining of the new subunits to other subunits already attached to the genome.
Figure~\ref{multinuc}b shows a snapshot of the simulations when the interaction is weak ($\epsilon_{hp}=-0.7$ $k_BT$), and the number of trimers is 40 with multiple nuclei already formed. We continue adding the subunits randomly to the edge of one of the existing subunits and accept the move as explained in the Methods section. The structure is still disordered. 

\begin{figure}

\centering
\includegraphics[width=0.47\textwidth]{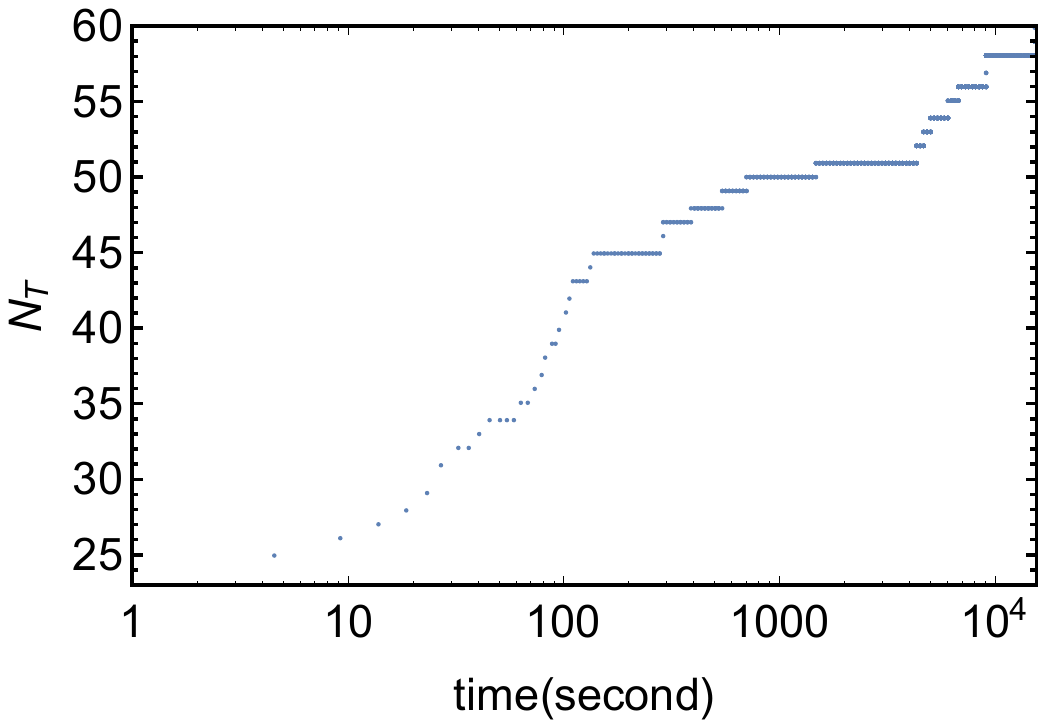}
\caption{Number of subunits assembled around the genome vs.~time (seconds) for multinucleation assembly at $\mu=-14k_BT$, $k_s$=600$k_BT$, $k_b$=200$k_BT$, and $\epsilon_{lj}=11.2k_BT$. The number of subunits increases from 23 to around 45 trimers in the first 130 s where the hydrophobic strength is weak ($\epsilon_{hp}=-0.7k_BT$). At $N_T=45$,  decreasing $\epsilon_{hp}$ to $-1k_BT$ the partially formed disordered capsid starts relaxing to the final structure with T=3 symmetry. The process of relaxation and reorganization to an ordered capsid takes around 14000 s. 
}
\label{timescalemulti}
\end{figure}

The previous self-assembly experiments \cite{Garmann2015,Chevreuil2018} showed that, when the pH is decreased, the attractive interaction between the protein subunits becomes stronger.  Following the experimental steps, we also decrease $\epsilon_{hp}$ from -0.7 $k_BT$ to -1 $k_BT$ when there are 45 trimers on the genome. At around 5000s, 54 trimers are relaxed to their minimum energy positions in  Fig.~\ref{multinuc}c. The shell will form a $T=3$ structure around 14000 s (Fig.~\ref{multinuc}d). The kinetics of assembly through the en masse pathway is illustrated in Fig.~\ref{timescalemulti} in the form of the number of subunits vs. time. The behavior of the growth curves obtained in the en masse assembly is very similar to those of Chevreuil {\it et al.} in that, upon decreasing pH, a disordered to ordered transition occurs. We have employed the results of Fig.~3a in ref.~\cite{Chevreuil2018} to calibrate the MC steps with time.

\begin{figure}
\centering
\includegraphics[width=1.0\linewidth]{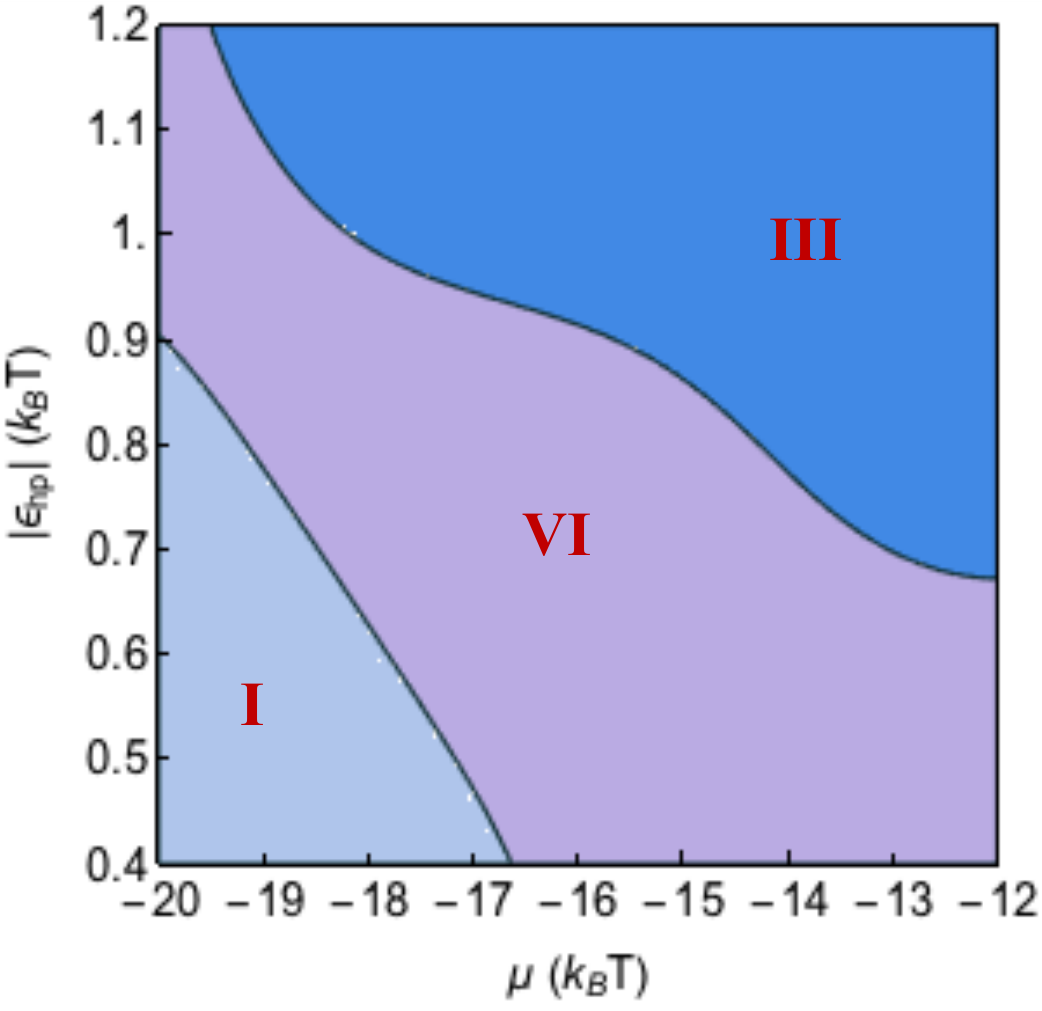}
\caption{Phase diagram of structures obtained from en masse simulations for different values of chemical potentials and the strength of hydrophobic interactions. The strong genome-protein attractive interaction ($\epsilon_{lj}=11.2$ $k_BT$) makes the assembly proceed through the en masse pathway. The blue shade corresponds to the region in which $T=3$ capsids form but the purple one represents the phase where amorphous structures assemble. No capsid nucleates in the light blue region. As we move from left to right, the concentration of free subunits increases and we observe the disordered to ordered transition. The other parameters in the simualtions are $k_s$=600$k_BT$, $k_b$=200$k_BT$, and the spontaneous radius $R_0=1.5$.} 
\label{phasewithcore}
\end{figure}

In addition to the pH change, the new experiments presented in the previous section reveal the important role of protein concentrations in the disordered to ordered transition. To explore the impact of protein concentration and protein-protein interaction and to explain the experiments presented in Fig.~\ref{fig:npc_fit_dammif}, we build a phase diagram for the formation of closed shells as a function of chemical potential and hydrophobic (attractive) interaction (see Fig.~\ref{phasewithcore}).  

Figure \ref{phasewithcore} shows that for fixed $k_s=600k_BT$ and $k_b=200k_BT$, at high protein concentrations and strong protein-protein attractive interaction, only $T=3$ capsids form (region III). Note that if we increase both protein concentration and the strength of protein interaction significantly, the structures will be stuck in a local minimum energy and aberrant particles form (not shown in the phase diagram).  In contrast at low hydrophobic interactions and low protein concentration, no structures nucleate (region I). In the purple region (VI) of the diagram, only messy/amorphous structures form as observed in the experiments at pH=7.5. In this regime, the protein-protein interaction is not strong enough to form an ordered structure. Nevertheless, the figure shows that for a fixed hydrophobic interaction, upon increasing protein concentrations, a disorder-order transition can occur; see for example the region around $\mu=-14k_BT$ and $\epsilon_{hp}=-0.8k_BT$. As illustrated in the figure, if for $\epsilon_{hp}=-0.8k_BT$, the chemical potential increases from -15 to -13 $k_BT$, one can observe the formation of $T=3$ structures, consistent with the new experimental results shown in Fig.~\ref{fig:npc_fit_dammif}.
It is worth mentioning that the transition from a disordered nucleoprotein complex to an icosahedral virion was reported several times by lowering the pH \cite{Chevreuil2018,Garmann2016,garmann2014assembly,cadena2012self}, which is equivalent to increasing $\epsilon_{hp}$, for a fixed $\mu$ in Fig.~\ref{phasewithcore}. The novelty of the experiments in this paper corresponds to the evidence of a disorder-order transition at a fixed $\epsilon_{hp}$ – through fixed pH and ionic strength – by decreasing $\mu$ {\it via} the increase of subunit concentration. Consistent with the phase diagram depicted in Fig.~\ref{phasewithcore}, SAXS measurements show that upon an increase of subunit concentration (the subunit-to-RNA mass ratio is set to 6 in all cases), the nucleoprotein complexes become gradually ordered and cryo-TEM reveals the presence of a number of well assembled virions at high concentration.

Quite interestingly, when we change the size of the core to be commensurate with a capsid, whose diameter is 23 $nm$ consistent with the assembly experiments in the presence of PSS, we obtain a structure with the symmetry of a tennis ball (see Appendix Fig.~19), which has been previously observed in self-assembly studies of clathrin shells \cite{wagner2015robust}. The phase diagram for PSS is very similar to that for RNA, but the boundary lines between ordered-disordered configurations have moved (see Fig. ~20 in Appendix). We find that capsid proteins assemble to form closed shells at lower protein concentration that are needed for assembly with RNA. This behavior supports at least qualitatively the experimental results with PSS \cite{Chevreuil2018}.

\section{\label{conclusion}Conclusion}
Since during the assembly of viral particles, the intermediate states are transient and as such not easily accessible experimentally, the most fundamental questions about assembly pathways have remained unresolved.  In this paper, we have focused on two different mechanisms observed in recent experiments:  one supporting the en masse assembly and the other the nucleation and growth pathway. 

\begin{figure}
\centering
\includegraphics[width=0.95\linewidth]{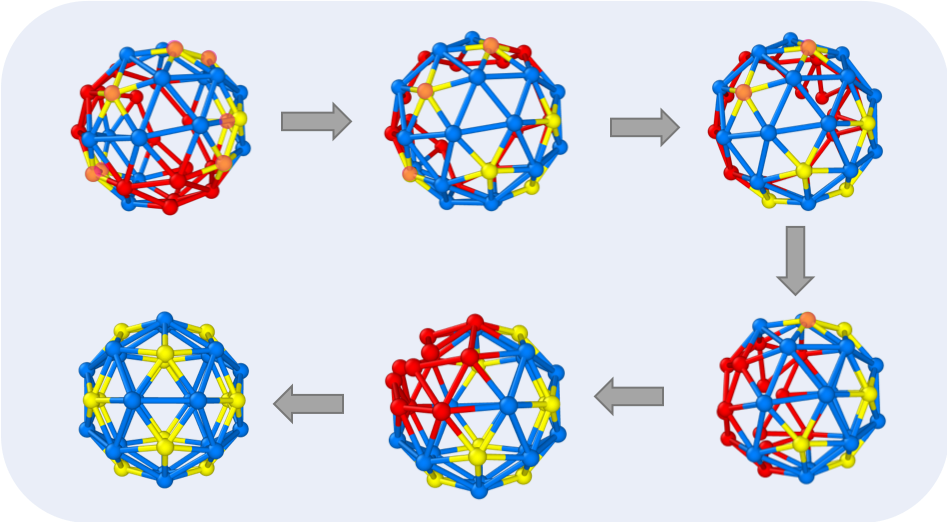}
\caption{Relaxation of a messy shell to an icosahedral structure. At the beginning, the shell has 57 trimers with 7 pentamers formed in ``wrong" positions (marked with orange color). The shell relaxes to a T=3 capsid upon a change in pH or protein concentration. The blue, yellow, and red colors correspond to hexamers, pentamers, and the edge respectively. The parameters used are stretching rigidity $k_s=200k_B T$, bending rigidity $k_b=200k_B T$, hydrophobic interaction $\epsilon_{hp}=1.0k_B T$, and chemical potential $\mu=-14k_B T$. The images are made with OVITO\cite{ovito}.}
\label{messygrowth2}
\end{figure}

\begin{figure}
\centering
\includegraphics[width=1\linewidth]{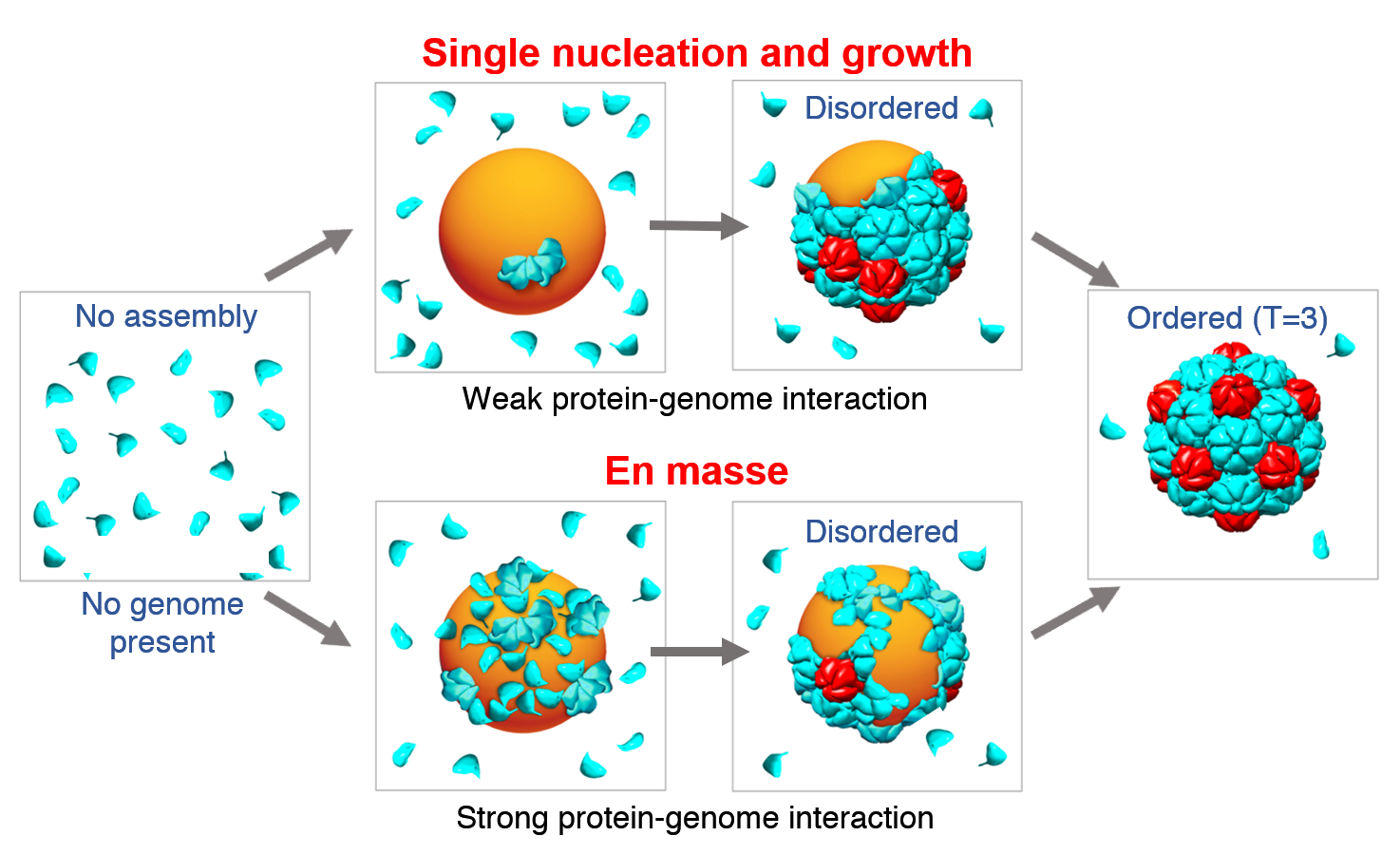}
\caption{The figure depicts the solution condition in which the protein concentration is not high enough for a capsid to nucleate in the absence of genome. In the presence of genome, depending on the strength of the attractive interaction between the genome (the orange spherical core) and proteins (blue subunits), the assembly proceeds either through the nucleation and growth or the en masse mechanism. In both cases, the intermediates can be disordered.  However, due to the strength of the elastic energy, at the end both shells become ordered, assuming a $T=3$ icosahedral structure. The red subunits belong to the pentameric defects.}
\label{cartoon}
\end{figure}

Our results show that the elastic energy plays a crucial role in the assembly of error-free symmetric shells in both cases. 
We have found that, as a shell grows, if the pentamers form in the positions that break icosahedral order,  through a reversible pathway, the pentamers can move to preserve the symmetry and form a $T=3$ shell. This is counterintuitive as many bonds need to break for a pentamer formed far from the edge, well inside the capsid, to move. However, our results show that the elastic energy becomes so strong that as many bonds as necessary can be broken to move the pentamer, see Fig.~\ref{penfix}.  Obviously, the relative strength of hydrophobic interaction and elastic energy are very important in this process, see Figs.~\ref{rev_phase} and \ref{fig:barrier}. Figure \ref{messygrowth2} shows a messy shell that was initially assembled fast in the presence of high protein concentration. These sorts of shells can form either through nucleation and growth or through an en masse pathway, see Fig.~\ref{cartoon}.  Upon decreasing the protein concentration, we find that the shell relaxes to a $T=3$ structure through a reversible pathway. The elastic energy indeed plays a crucial role in preserving the symmetry in transferring disordered to ordered structures.

\begin{figure}
\centering
\includegraphics[width=1.03\linewidth]{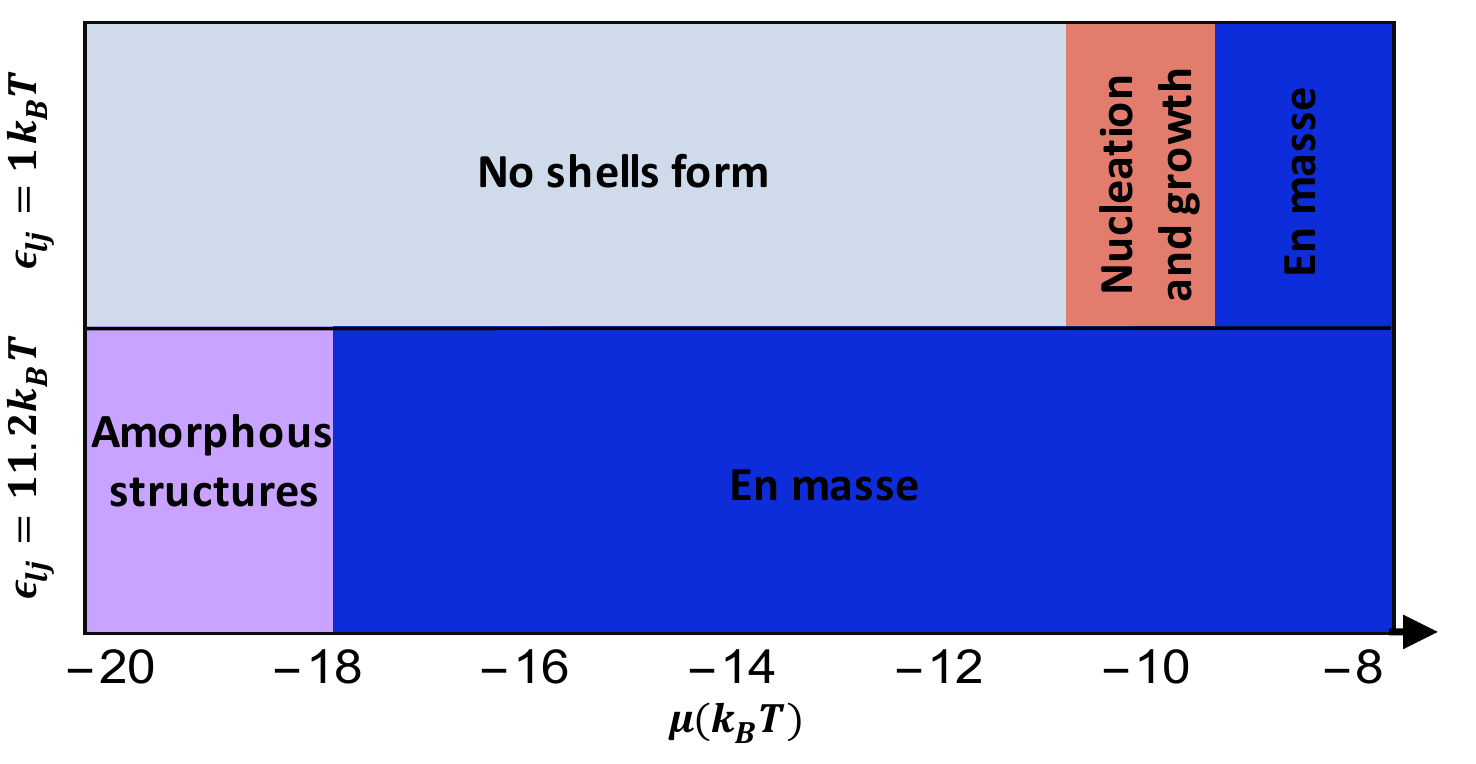}
\caption{Capsids form with different mechanisms based on the strength of the protein-genome interaction. At a weak Lennard-Jones interaction ($\epsilon_{lj}=1.0 k_BT$), for the chemical potentials between -11 $k_BT$ and -9 $k_BT$, capsids form through the nucleation and growth mechanism.  For higher chemical potentials, the en masse assembly becomes the dominant pathway for the formation of $T=3$ structures. However, at strong Lennard-Jones interaction ($\epsilon_{lj}=11.2$ $k_BT$) we obtain amorphous structures for low chemical potentials, but T=3 structures form through the en masse pathway for higher chemical potentials. Other parameters used in the simulations are $k_s$=600$k_BT$, $k_b$=200$k_BT$, and $\epsilon_{hp}=-1$ $k_BT$.}
\label{enmasseandnucl}
\end{figure}

The important role of protein concentration in the assembly pathway is summarized in Fig.~\ref{enmasseandnucl} for two different values of genome-protein interaction, $\epsilon_{lj}$. If the genome-protein interaction is weak  $\epsilon_{lj}=1 k_BT$, for the low protein concentration, no shell nucleates. As the protein concentration increases, the shell grows through the nucleation and growth pathway.  At higher concentrations, we observe that subunits get adsorbed onto the spherical core and shell grows through en masse assembly into a $T=3$ structure. In contrast for large $\epsilon_{lj}=11.2 k_BT$ at low protein concentration, amorphous structures form. Upon increasing protein concentration ordered structures are obtained through en masse assembly.  

 These results are consistent with our new SAXS measurements, which clearly show an increasing degree of order upon the increase of concentration.  While even our best curves are not as regular as those obtained with native virions, in marked contrast with the previous experiments we find that some structures look spherical and fully assembled. We note that the concentration may not be high enough or other factors such as crowding effect or divalent cations may be important in the formation of icosahedral capsids as well. Nevertheless, our experimental and numerical results clearly demonstrate that the chemical potential can significantly lower the energy barrier between amorphous complexes and virions, and it must have definitively a role \textit{in vivo} assembly as well.


Deciphering the factors that contribute to the dominant assembly pathway could enable biomedical attempts to block viral replication and infection and promote the use of capsids in bionanotechnolgy. 

\begin{acknowledgments}
This work was supported by the NSF through Awards DMR-1719550.  The plasmid coding for the CCMV RNA 2 was kindly provided by Christian Beren from William Gelbart's group at the University of California Los Angeles. G.T. acknowledges financial support from the Agence Nationale de la Recherche (Contract ANR-16-CE30-0017-01). The electron microscopy imaging is supported by ``Investissements d'Avenir" LabEx PALM (Contract ANR-10-LABX-0039-PALM), and we thank J\'eril Degrouard for his technical assistance with the microscope. L.M., R.L.R., and G.T. also acknowledge the European Synchrotron Radiation Facility (ESRF) and the SOLEIL synchrotron for allocation of synchrotron beam time on ID02 and SWING beamlines.
\end{acknowledgments}

\appendix*
\section{}

\subsection{Elastic energy of a spherical cap}
\begin{figure}
\centering
\includegraphics[width=1\linewidth]{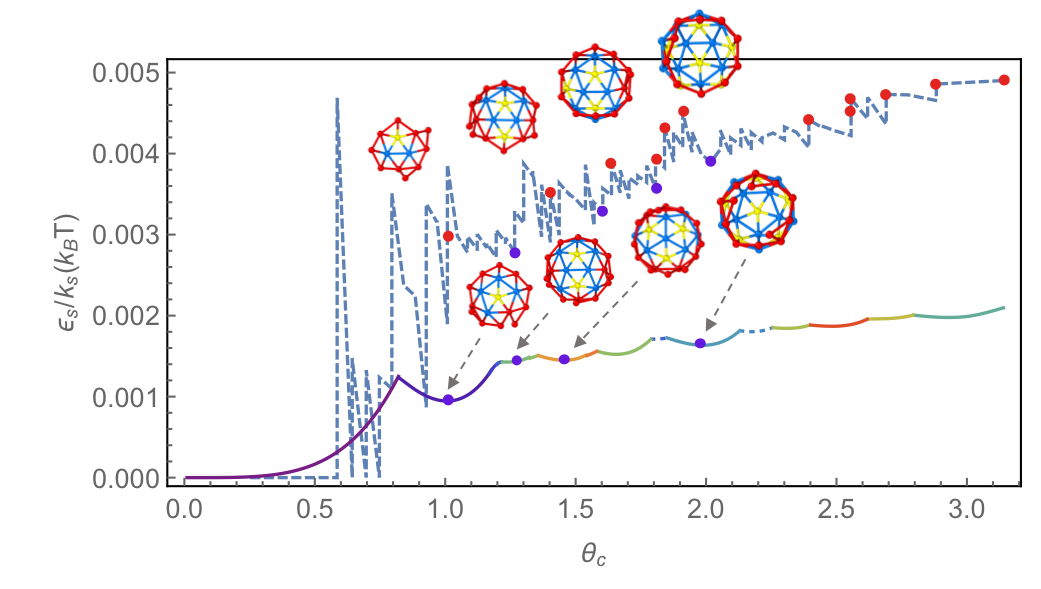}
\caption{The elastic energy of a spherical cap as a function of $\theta_c$, which is related to the cap $area=2\pi R^2(1-\cos\theta_c)$. The formation of each new pentamer is marked with a red dot. Some snapshots of simulations are shown above the dashed line. Each pentamer first forms at the cap boundary and then the shell grows around it. After relaxation, the pentamers have moved towards the center of the cap. The relaxed configurations are shown below the dashed line and are marked with purple dots. The solid line shows the ground state energy obtained through continuum elastic theory, and the purple dots illustrate the same configurations as the ones found in simulations. There is a perfect match between the structures obtained in the simulations and continuum theory.
The parameters used for the simulations are the spring modulus $k_s=200k_B T$, bending rigidity $k_b=200k_B T$, hydrophobic interaction $\epsilon_{hp}=-1.6k_B T$ and chemical potential $\mu=-14.6k_B T$. The final structure has T=3 icosahedral symmetry.}
\label{fixpen1}
\end{figure}

The elastic energy of a spherical cap as a function of $\theta_c$, which is related to the cap area is shown in Fig.~\ref{fixpen1}.

\subsection{Disordered to ordered transition and the role of elastic energy}
\begin{figure*}
\centering
\includegraphics[width=1\textwidth]{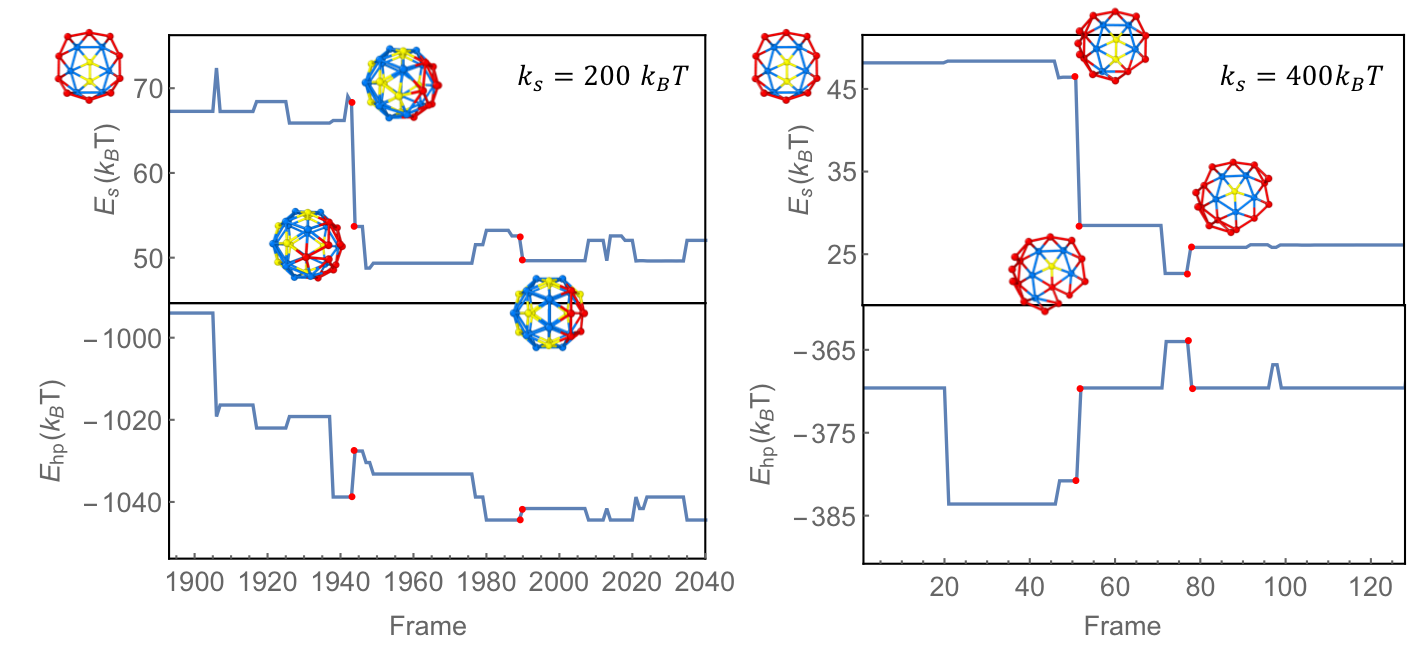}
\caption{Two examples of disorder-order transitions with different spring constants $k_s=200k_B T$ and $400k_B T$. The distribution of pentamers in ordered structures possesses subgroup symmetries of icosahedron.  The red dots illustrate the step in which a pentamer formed in a ``wrong" position dissociates and becomes a hexamer. For $k_s=200k_B T$, this occurs after $1990$ MC steps when the capsid has $56$ trimers. The higher elastic energy, the faster pentamers transforms into hexamers if formed in ``wrong" positions.  For $k_s=400k_B T$, it takes $78$ steps for a pentamer to move and ``correct" its location such that the cap has only 25 trimers when the pentamer formed in the ``wrong" position dissociates. We note that each MC move involves any attempt to move any trimer or vertex through diffusion, growth or detachment. The simulations are performed for the bending rigidity $k_b=200k_B T$, hydrophobic interaction $\epsilon_{hp}=-1.4k_B T$, and chemical potential $\mu=-14.6k_B T$.}
\label{fixpen}
\end{figure*}

Two examples of disorder-order transitions with different spring constants are shown in Fig.~\ref{fixpen}

\subsection{Hydrophobic interaction}
The total hydrophobic energy of a capsid is
\begin{equation}
E_{hp}=\sum_{v_i} \epsilon_{hp}[nt_{v_i} \cdot (nt_{v_i}-1)],\label{Ehp}
\end{equation}
with $v_i$ the vertex index, $\epsilon_{hp}$ the strength of the monomer-monomer interaction and $nt_{v_i}$ the number of triangles sharing the vertex $v_i$.
For a complete shell with $Q$ triangles, the total number vertices is $N_v=0.5Q+2$ with twelve pentamers and the rest hexamers. The total hydrophobic energy of a complete shell based on Eq.~\ref{Ehp} can thus be written as $E_{hp}=12\epsilon_{hp}\times20+(0.5Q+2-12)\epsilon_{hp}\times30=15\epsilon_{hp}(Q-4)$. The hydrophobic interaction per trimer subunit is then equal to $\epsilon=15\epsilon_{hp}\frac{Q-4}{Q}$. For a T=3 capsid with $Q=60$ trimers, $\epsilon=14\epsilon_{hp}$. When $Q\rightarrow\infty$, almost every triangle is part of a hexamer with 15 monomer-monomer interactions, and as such the hydrophobic interaction then converges to $15\epsilon_{hp}$ per subunit.

\begin{figure}[h]
    \centering
    \includegraphics[width=1.0\linewidth]{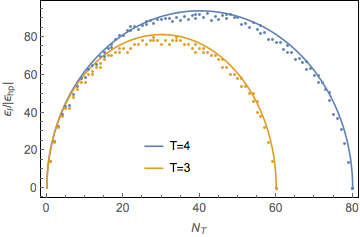}
    \caption{Plots of line tensions vs. number of subunits $N_T$. The figure compares the line tension $\epsilon_l$ obtained from Eq.~\ref{eqEhp} (solid lines) and the results of simulations (dots) for T=3 ($Q=60$) and T=4 ($Q=80$) capsids. $T=4$ structure has higher line tension.}
    \label{fig:linetension}
\end{figure}

For an incomplete shell, the vertices are categorized as the pentameric vertex $v_5$, the hexameric vertex $v_6$ and the edge vertex $v_e$. The total number of trimers $N_T$ then statisfy the constraint $3N_T=5v_5+6v_6+3v_e$.
Following the results of Ref.\cite{li2018large,lizandigrason}, we assume that pentamers grow linearly with the capsid area, so the pentameric vertex number is $v_5=\frac{12N_T}{Q}$.
The edge vertex number $v_e$ and the perimeter $l_p$ are related through $v_e=\frac{l_p}{l_0}$ with $l_p$ equal to
\begin{eqnarray}\label{eqlp}
l_p&=&2\pi R \sin{\theta_c} \nonumber\\
&=&2\pi R\sqrt{1-(1-A/(2\pi R^2))^2} \nonumber\\
&=&2\sqrt{\pi}\sqrt{A-A^2/4\pi R^2} \nonumber\\
&=&2\sqrt{\pi}\sqrt{N_T a-(N_T a)^2/(Q a)} \nonumber\\
&=&2\sqrt{\pi}\sqrt{a}\sqrt{\frac{N_T}{Q}(Q-N_T)},
\end{eqnarray}
and $a=\frac{\sqrt{3}}{4}l_0^2$ the area of one trimer. $l_0$ is the length of each trimer side.
The number of edge vertices and hexameric vertices then becomes $v_e=\sqrt{\sqrt{3}\pi}\sqrt{\frac{N_T}{Q}(Q-N_T)}$ and  $v_6=\frac{1}{6}(3N_T-5v_5-3v_e)$, respectively. Finally we can obtain the hydrophobic energy of a partially formed shell by assuming each vertex at the edge on average has three neighbors.  The total hydrophobic energy of an incomplete shell will then be
\begin{eqnarray}\label{eqEhp}
E_{hp}&=&\epsilon_{hp} (5\times4v_5+6\times5v_6+3\times2v_e) \nonumber\\
&=&\epsilon_{hp} \left(20v_5+5(3N_T-5v_5-3v_e)+6v_e\right) \nonumber\\
&=&\epsilon_{hp} (15N_T-5v_5-9v_e) \nonumber\\
&=&\epsilon_{hp} \left(15N_T-\frac{60N_T}{Q}-9\sqrt{\sqrt{3}\pi}\sqrt{\frac{N_T}{Q}(Q-N_T)}\right) \nonumber\\
&=&15\epsilon_{hp} \frac{Q- 4}{Q}N_T-15\epsilon_{hp}\sqrt{\alpha}\sqrt{\frac{N_T}{Q}(Q-N_T)} \nonumber\\
&=&\epsilon N_T + \epsilon_{l},
\end{eqnarray}
with $\alpha=\frac{9\sqrt{3}\pi}{25}$ and $\epsilon=15\epsilon_{hp} \frac{Q-4}{Q}$. The quantity $\epsilon_l$ is called the line tension in classical nucleation theory. The comparison of the line tension calculated through Eq.~\ref{eqEhp} and obtained through our simulations is illustrated in Fig.~\ref{fig:linetension} for $T=3$ and $T=4$ capsids with $Q=60$ and $80$. We find a very good match between the results of the simulations (dotted lines) and the simple theory (Eq.~\ref{eqEhp}) presented above (dashed lines).

\subsection{Detailed protocols for sample preparation}

70 g of infected leaf materials were blended with 150 mL of 0.15 M sodium acetate pH 4.8 and 150 mL of ice-cold chloroform. After centrifugation at 10,000$\times\ g$ for 10 min, the virions in the supernatant were precipitated by adding NaCl to a final concentration of 20 mM along with 8\% (v.w$^{-1}$) poly(ethylene glycol) (MW 8000). The solution was centrifuged at 10,000$\times\ g$ for 10 min and the pellet was resuspended in 50 mM sodium acetate pH 4.8. After centrifugation at 8,000$\times\ g$ for 10 min, the supernatant containing virions was ultracentrifuged through a 20\% (v.w$^{-1}$) sucrose cushion in water at 150,000$\times\ g$ for 2 h, and the virions in the pellet were stored at -80 $^\circ$C until use. 

For protein purification, 5 mg of virions were dialyzed against 50 mM Tris-HCl pH 7.5, 0.5 CaCl$_2$, 1 mM ethylenediaminetetraacetic acid (EDTA) pH 8.0, 1 mM phenylmethylsulfonyl fluoride (PMSF), 1 mM dithiothreitol (DTT), and centrifuged at 150,000$\times\ g$ for 18 h. Proteins were collected in the supernatant and stored in 50 mM sodium acetate pH 4.8, 0.5 M NaCl, 1 mM EDTA pH 8.0 at 4 $^\circ$C until use. Full viral RNA genome comprising the four RNA segments encoding for the complete CCMV genome was purified by mixing 0.1 mL of virions with 1 mL of TRIzol$\circledR$ reagent (Life Technologies, France). After addition of 0.2 mL of chloroform, the solution was centrifuged at 12,000$\times\ g$ for 15 min and the aqueous phase was removed prior to the addition of 0.5 mL of isopropanol. After centrifugation at 12,000$\times\ g$ for 15 min, the pellet was washed with 1 mL of 75\% ethanol and centrifuged at 7,500$\times\ g$ for 5 min. The RNA pellet was resuspended in RNase-free water and stored at -80 $^\circ$C until use. CCMV RNA 2 was transcribed \textit{in vitro} as follows: The plasmid coding for the RNA was transformed in NEB$\circledR$ 10-$\beta$ competent \textit{E. coli} cells (New England Biolabs, MA, USA), produced and purified by a NucleoBond® Xtra plasmid DNA purification kit (Macherey-Nagel, Germany), and linearized using the XbaI restriction enzyme. The linearized plasmid containing a T7 promoter was then transcribed using a MEGAscript$\circledR$ Kit (Thermo Fisher Scientific, MA, USA). Newly produced RNAs were finally purified using a MEGAclear$\circledR$ Transcription Clean-Up Kit (Thermo Fisher Scientific, MA, USA). Purity was checked by running an agarose gel electrophoresis and RNA was stored in RNase-free water at -80 $^\circ$C. Capsid proteins and RNA were checked by spectrophotometry, and verified $A_\mathrm{280}/A_\mathrm{260}>1.65$ and $A_\mathrm{260}/A_\mathrm{280}>1.8$, respectively.

Nucleoprotein complexes (NPCs) were assembled as follows: Capsid proteins were initially dialyzed against capsid disassembly buffer, i.e., 50 mM Tris-HCl pH 7.5, 450 mM NaCl, 1 mM EDTA pH 8.0, 1 mM PMSF and 1 mM DTT. For the measurements with the full genome, capsid proteins and RNA were separately dialyzed against NPC buffer, i.e., 50 mM Tris-HCl pH 7.5, 50 mM NaCl and 1 mM EDTA pH 8.0, and rapidly mixed together with a stopped-flow device at the desired final concentrations. For the measurements with \textit{in vitro} transcribed CCMV RNA 2, RNA was dialyzed against capsid disassembly buffer, then mixed with capsid proteins at the desired final concentration, and the mixture was dialyzed against NPC buffer overnight.

\subsection{Polydisperse vesicle model}

\begin{figure}[h]
\includegraphics[width=1\linewidth]{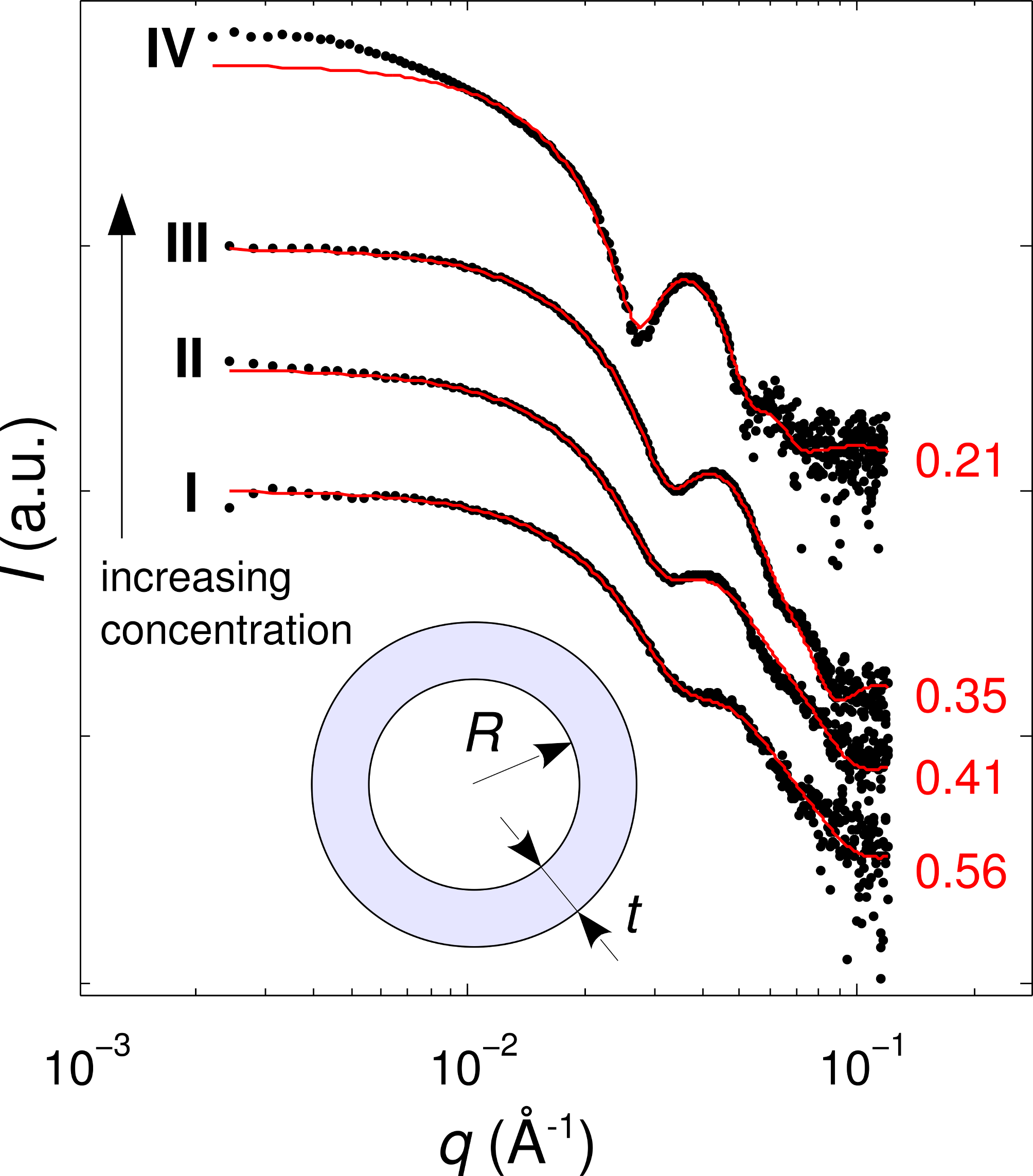}
\caption{\label{fig:npc_fit_vesicle} SAXS patterns of NPCs (black dots) fitted with a polydisperse vesicle model (red line). The model consists of homogeneous vesicles (see the drawing) with a fixed shell thickness $t$ and radii distributed normally about $R$ with a standard deviation $\Delta R$. The numbers in red next to the curves are the polydispersity indexes $\Delta R/R$. NPCs are made of the full RNA genome (\textbf{I}, \textbf{II} and \textbf{III}) with subunit concentrations of 0.5, 1.0 and 2.1 g.L$^{-1}$, respectively, and of \textit{in vitro} transcribed RNA 2 (\textbf{IV}) at a subunit concentration of 2.1 g.L$^{-1}$. In all cases, the subunit-to-RNA mass ratio is about 6. The scattering curves are shifted for clarity. The fitting parameters are given in Table~\ref{tab:fit_vesicle_parameters}.}
\end{figure}

Figure~\ref{fig:npc_fit_vesicle} depicts the scattering patterns of various NPC samples fitted with a polydisperse vesicle model. This model is not intended to represent realistically the structure of the NPCs but rather to give a rough measure of their size and shape uniformity. The fitted patterns of samples \textbf{I}, \textbf{II} and \textbf{III}, made up with the full CCMV genome, show that upon increasing the subunit concentration from 0.5 to 2.1 g.L$^{-1}$, the polydispersity index $\Delta R/R$ decreases from 0.56 to 0.35, which suggests an increasing degree of order in the structure of the NPCs, in good agreement with the simulated disorder-order transition. Note that the subunit-to-RNA mass ratio is set to 6, that is, well larger than the value of 4 found in native virions; in other words, there are in principle always more than enough subunits to build up a closed capsid with genomic RNA inside. For NPCs made with only \textit{in vitro} transcribed RNA 2 (sample \textbf{IV}), the degree of order looks even higher since the polydispersity index drops to 0.21, which can be ascribed to a lesser variability of the NPC structure than in the presence of multiple RNA segments.

\begin{table*}[ht]
\caption{\label{tab:fit_vesicle_parameters}Sample and fitting parameters for the SAXS patterns on Fig.~\ref{fig:npc_fit_vesicle} using a polydisperse vesicle model.}

\begin{ruledtabular}
\begin{tabular}{lccccc}
Sample & Subunit concentration (g.L$^{-1}$) & Subunit-to-RNA mass ratio & $R$ (nm) & $\Delta R/R$ & $t$ (nm) \\
\colrule
\textbf{I} & 0.5 & 6 & 40 & 0.56 & 57\\
\textbf{II} & 1.0 & 6 & 52 & 0.41 & 60\\
\textbf{III} & 2.1 & 6 & 46 & 0.35 & 74\\
\textbf{IV} & 2.1 & 6 & 62 & 0.21 & 85
\end{tabular}
\end{ruledtabular}
\end{table*}

\subsection{Nanotubes}

\begin{figure}
\includegraphics[width=1\linewidth]{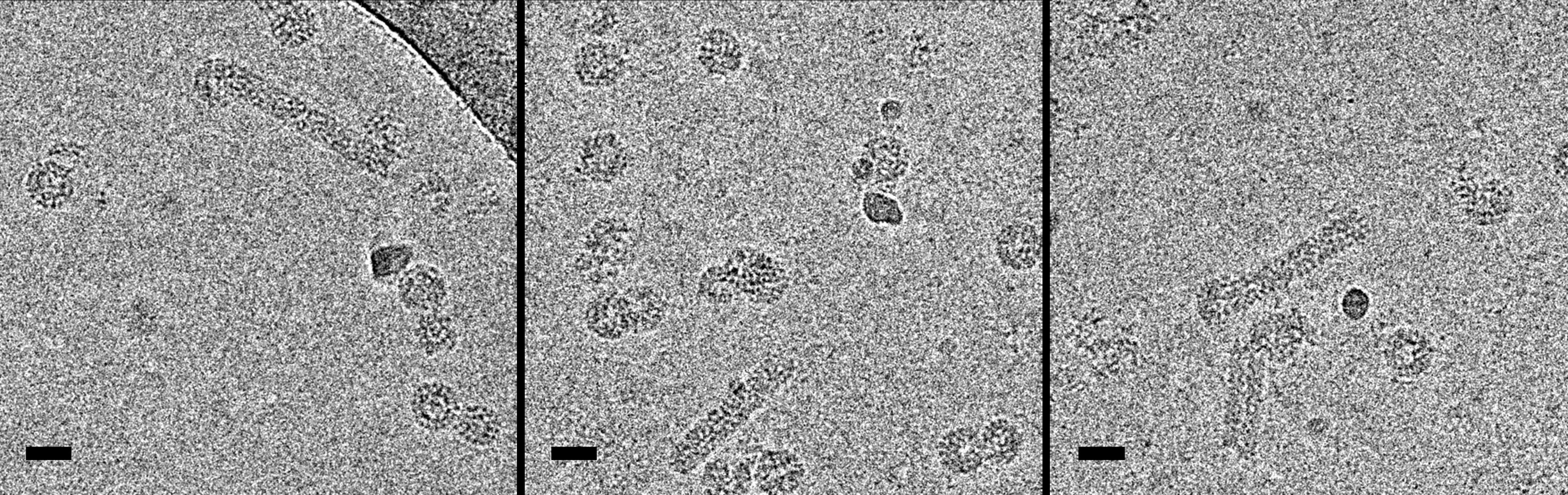}
\caption{\label{fig:nanotubes_cryotem} Cryo-transmission electron microscopy images of nanotubes coexisting with spherical NPCs at neutral pH. The sample contains 3.9 g.L$^{-1}$ of subunits along with \textit{in vitro} transcribed RNA 2 at a subunit-to-RNA mass ratio of 6. Scale bars are 30 nm.}
\end{figure}

Nanotubes are seen at high concentrations of subunits and \textit{in vitro} transcribed RNA 2 (Fig.~\ref{fig:nanotubes_cryotem}). These nanotubes are probably formed by the excess of free subunits that do not bind on RNA. The presence of nanotubes at neutral pH is consistent with the phase diagram of CCMV CPs \cite{Lavelle2009}.

\subsection{Simulations with smaller cores}

Many experiments show that the capsids assembled around PSS are smaller \cite{Chevreuil2018}. To this end, we construct a phase diagram (Fig.~\ref{smallerphase}) as a function of $\epsilon_{hp}$ and $\mu$, similar to the one in Fig. 11 in the text. We use en masse simulations with small cores($R_0=1.2$), comparable to the capsids filled with PSS. The smaller structure looks like a tennis ball, previously observed in the self-assembly studies of clathrin shells \cite{fotin2004molecular,wagner2015robust}.

\begin{figure}[h]
\centering
\includegraphics[width=0.5\textwidth]{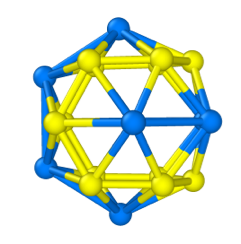}
\caption{ The structure formed in the simulations when the size of the core is $R_0=1.2$. This structure looks like a tennis ball and has been observed in the clathrin shell experiments too \cite{fotin2004molecular,wagner2015robust}.
}
\label{tennisball}
\end{figure}

\begin{figure}
\centering
\includegraphics[width=0.5\textwidth]{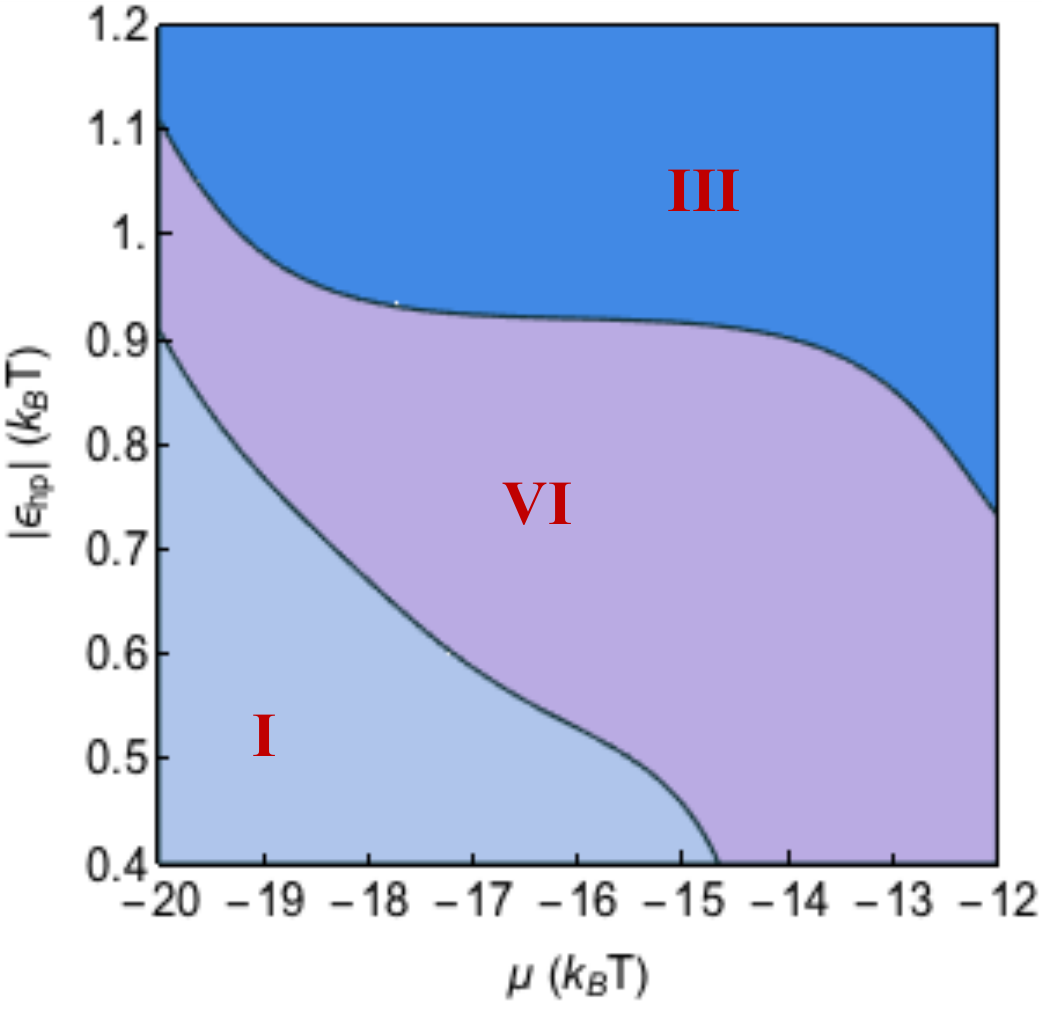}
\caption{Phase diagram of structures obtained from en masse simulations in the presence of smaller core size comparing to Fig. 11 in the paper. The blue shade corresponds to the region in which closed capsids, mostly with tennis ball symmetry form but the purple one represents the phase where amorphous structures form. The light blue region corresponds to region where no shell forms.  Parameters that are used in this simulations are $R_0=1.2$, $\epsilon_{lj}=11.2 k_BT$, $k_s$=600$k_BT$ and $k_b$=200$k_BT$. }
\label{smallerphase}
\end{figure}

\pagebreak
\bibliography{apssamp}


----------------------------------------------




\end{document}